\documentclass[12pt]{iopart}
\usepackage{iopams}
\usepackage{setstack}
\usepackage{amssymb}
\usepackage{latexsym}
\usepackage{color}
\usepackage{graphicx}   
     
\begin{document}
\title[Finite oscillator]{Finite oscillator obtained through\\ finite frame quantization}
\author{Nicolae Cotfas and Daniela Dragoman}
\address{University of Bucharest,  Physics Department,\\ P.O. Box MG-11, 077125 Bucharest, Romania}

\eads{\mailto{ncotfas@yahoo.com}, \mailto{danieladragoman@yahoo.com}}

\begin{abstract}
The Hamiltonian of the harmonic oscillator is usually defined as a differential operator, 
but an integral representation can be obtained by using the coherent state quantization.
The finite frame quantization is a finite counterpart of the coherent state quantization and
it allows us to define a finite oscillator by starting from the integral representation of the harmonic oscillator.
Our purpose is to investigate the oscillator obtained in this way, and to present a possible application to the discrete fractional Fourier transform.
\end{abstract}

\section{Introduction}

The harmonic oscillator plays a fundamental role in quantum mechanics. A finite-dimensional version, leading to Harper functions, can be obtained in a natural way by using a finite difference operator instead of the differential operator and the Fourier invariance \cite{Ba}. Despite the fact that the eigenvalues and eigenfunctions can be obtained only numerically, the finite-dimensional quantum system obtained in this way has some important applications. For example, the version of the discrete fractional Fourier transform based on it \cite{Ca}  is used in optics and signal processing \cite{OZK}.

The finite frame quantization \cite{CG}, which is a finite counterpart of the quantum state quantization \cite{Ga}, allows us to define an alternative finite-dimensional version of the harmonic oscillator \cite{CGV}. Our main purpose is to investigate mathematically and numerically  the quantum system with finite-dimensional Hilbert space obtained in this way, and to present an application to the discrete fractional Fourier transform.

\section{Fourier transform and the finite Fourier transform}
The Fourier transform of a function $\psi \!:\!\mathbb{R}\!\longrightarrow \!\mathbb{C}$ belonging to 
$L^1(\mathbb{R})\cap L^2(\mathbb{R})$ is 
\begin{equation}
\mathcal{F}[\psi ]:\mathbb{R}\longrightarrow \mathbb{C},\qquad  \mathcal{F}[\psi ](x)
=\frac{1}{\sqrt{2\pi }}\int_{-\infty }^\infty {\rm e}^{-{\rm i}xx'}\psi (x')\, dx'.
\end{equation}
This transformation can be extended to the Fourier-Plancherel transform on $L^2(\mathbb{R})$. It is a unitary transformation,
\begin{equation}
\mathcal{F}^{-1}[\psi ](x)=\frac{1}{\sqrt{2\pi }}\int_{-\infty }^\infty {\rm e}^{{\rm i}xx'}\psi (x')\, dx'
\end{equation}
and satisfies the relations
\begin{equation}
 \mathcal{F}^2[\psi ](x)=\psi (-x)\, ,\qquad \mathcal{F}^4[\psi ](x)=\psi (x)\, ,\qquad \mbox{for\ any}\quad x\!\in \!\mathbb{R}.
\end{equation}
In order to obtain a finite counterpart,  we consider an odd integer $d\!=\!2s\!+\!1$ and the set
\[
\mathcal{R}_d\!=\!\left\{ -s\sqrt{\delta},\, (-s\!+\!1)\sqrt{\delta},\, \dots \, , 
\, (s\!-\!1)\sqrt{\delta},\, s\sqrt{\delta}\right\}\qquad \mbox{with}\quad \delta \!=\!\frac{2\pi }{d}.
\]
Each function
\[
\varphi :\mathbb{Z}_d\sqrt{\delta}\longrightarrow \mathbb{C}
\]
that is, each  periodic function
\[
\varphi :\mathbb{Z}\sqrt{\delta}\longrightarrow \mathbb{C}
\]
satisfying the relation
\begin{equation}
\varphi (n\sqrt{\delta})=\varphi (n\sqrt{\delta}+d\sqrt{\delta}),\qquad \mbox{for any}\qquad n\!\in \!\mathbb{Z}
\end{equation}
 is well-determined by its restriction to the set $\mathcal{R}_d$.
The space $l^2(\mathcal{R}_d)$ of all the functions $\varphi :\mathbb{Z}_d\sqrt{\delta}\longrightarrow \mathbb{C}$  considered with the scalar product
\[
\begin{array}{l}
\langle \varphi _1,\varphi _2\rangle =\sum\limits_{n=-s}^s\overline{\varphi _1(n\sqrt{\delta})}\,\varphi _2(n\sqrt{\delta})
\end{array}
\]
is a Hilbert space isomorphic to the standard $d$-dimensional Hilbert space $\mathbb{C}^d$.
Since
\[
\lim_{d\rightarrow \infty }\sqrt{\delta}=0\qquad \mbox{and} \qquad \lim_{d\rightarrow \infty }(\pm s)\sqrt{\delta}=\pm \infty
\]
we can consider that, in a certain sense,
\begin{equation}
\mathcal{R}_d\stackrel{\scriptstyle{d\rightarrow \infty }}{-\!\!-\!\!\!\longrightarrow }\mathbb{R}\qquad \mbox{and} \qquad  l^2(\mathcal{R}_d)\stackrel{\scriptstyle{d\rightarrow \infty }}{-\!\!-\!\!\!\longrightarrow }L^2(\mathbb{R}).
\end{equation}

The finite Fourier transform of a function $\varphi :\mathbb{Z}\sqrt{\delta}\longrightarrow \mathbb{C}$ from $l^2(\mathcal{R}_d)$ is the function 
\begin{equation}
\begin{array}{l}
{\bf F}[\varphi ] :\mathbb{Z}\sqrt{\delta}\longrightarrow \mathbb{C},\qquad 
{\bf F}[\varphi ](u)=\frac{1}{\sqrt{d}}\sum\limits_{v \in \mathcal{R}_d}{\rm e}^{-{\rm i}uv}\, \varphi (v)
\end{array}
\end{equation}
that is, ${\bf F}$ is the transformation 
$l^2(\mathcal{R}_d)\longrightarrow l^2(\mathcal{R}_d):\, \varphi \mapsto {\bf F}[\varphi ]$, where
\begin{equation}
\begin{array}{l}
{\bf F}[\varphi ](n\sqrt{\delta})=\frac{1}{\sqrt{d}}\sum\limits_{k=-s}^s{\rm e}^{-\frac{2\pi {\rm i}}{d}nk}\, \varphi (k\sqrt{\delta}).
\end{array}
\end{equation}
The inverse of ${\bf F}$ is the adjoint transformation $l^2(\mathcal{R}_d)\!\rightarrow \!l^2(\mathcal{R}_d):\, \varphi \!\mapsto \!{\bf F}^+[\varphi ]$, defined by
\begin{equation}
\begin{array}{l}
{\bf F}^+[\varphi ](n\sqrt{\delta})=\frac{1}{\sqrt{d}}\sum\limits_{k=-s}^s{\rm e}^{\frac{2\pi {\rm i}}{d}nk}\, \varphi (k\sqrt{\delta}).
\end{array}
\end{equation}
This means that ${\bf F}^+{\bf F}={\bf F}{\bf F}^+={\bf I}$, where ${\bf I}$ is the identity operator ${\bf I}\varphi =\varphi $.
The finite Fourier transform satisfies the relations
\begin{equation}
{\bf F}^2[\varphi ](u)=\varphi (-u),\qquad {\bf F}^4[\varphi ](u)=\varphi (u).
\end{equation}
For $d\!> \!4$, the eigenvalues of $\mathbf{F}$ are $1$, $-{\rm i}$, $-1$, ${\rm i}$. The operators 
\begin{equation}
\begin{array}{ll}
\pi _0=\frac{1}{4}(\mathbf{I}+\mathbf{F}+\mathbf{F}^2+\mathbf{F}^3)\qquad & \pi _2=\frac{1}{4}(\mathbf{I}-
\mathbf{F}+\mathbf{F}^2-\mathbf{F}^3)\\[2mm]
\pi _1=\frac{1}{4}(\mathbf{I}\!+\!{\rm i}\mathbf{F}\!-\!\mathbf{F}^2-{\rm i}\mathbf{F}^3)\qquad & 
\pi _3=\frac{1}{4}(\mathbf{I}\!-\!{\rm i}\mathbf{F}-\mathbf{F}^2\!+\!{\rm i}\mathbf{F}^3)
\end{array}
\end{equation}
are the corresponding orthogonal projectors, and $\mathbf{F}$ admits the spectral representation
\begin{equation}
\mathbf{F}=\pi _0-{\rm i}\pi _1-\pi _2+{\rm i}\pi _3=\sum_{m=0}^3(-{\rm i})^m\, \pi _m
=\sum_{m=0}^3{\rm e}^{-\frac{\pi {\rm i}}{2}m}\, \pi _m.
\end{equation}

The relation
\begin{equation}\label{periodicdelta}
\sum_{a=-s}^s{\rm e}^{\frac{2\pi {\rm i}}{d}a x}={\rm e}^{-\frac{2\pi {\rm i}}{d}s x}\frac{{\rm e}^{2\pi {\rm i} x}-1}{{\rm e}^{\frac{2\pi {\rm i}}{d}x}-1}\qquad \mbox{for any}\quad x\!\in \!\mathbb{R}\!-\!d\mathbb{Z}
\end{equation}
allows us to obtain the equality
\begin{equation}
\sum_{a=-s}^s{\rm e}^{\frac{2\pi {\rm i}}{d}a n}=\left\{
\begin{array}{lll}
d & \mbox{for} & n\in d\mathbb{Z}\\
0 & \mbox{for} & n\not\in d\mathbb{Z}
\end{array} \right. 
\end{equation}
and to compute the finite Fourier transform of the coordinate function 
\[
\mathfrak{q} :\mathcal{R}_d\longrightarrow \mathbb{R},\qquad \mathfrak{q}(\alpha )=\alpha 
\]
and of its square
\begin{equation}
\begin{array}{l}
{\bf F}[\mathfrak{q}](n\sqrt{\delta })\!=\!\frac{\sqrt{2\pi }}{d}\sum\limits_{a=-s}^s a\, {\rm e}^{-\frac{2\pi {\rm i}}{d}a n}\!=\!\left\{
\begin{array}{cll}
0 & \mbox{for} & n \in d\mathbb{Z}\\[2mm]
(-1)^{n} \frac{{\rm i}\sqrt{\pi }}{\sqrt{2} \sin \frac{\pi }{d}n}& \mbox{for} & n\not\in d\mathbb{Z}
\end{array} \right. 
\end{array}
\end{equation}
\begin{equation}
\begin{array}{l}
{\bf F}[\mathfrak{q}^2](n\sqrt{\delta })\!=\!\frac{2\pi }{d\sqrt{d}}\sum\limits_{a=-s}^s a^2\,{\rm e}^{-\frac{2\pi {\rm i}}{d}a n}\!=\!\left\{
\begin{array}{cll}
\frac{2\pi }{\sqrt{d}}\, \frac{s(s+1)}{3} & \mbox{for} & n\in d\mathbb{Z}\\[2mm]
(-1)^n \frac{\pi \, \cos \frac{\pi }{d}n}{\sqrt{d}\, \sin ^2\frac{\pi }{d}n} & \mbox{for} & n\not\in d\mathbb{Z}.
\end{array} \right. 
\end{array}
\end{equation}

\section{Heisenberg-Weyl group and a finite counterpart}
A  quantum-mechanical system with one degree of freedom can be described by using the coordinate operator 
$\hat q$ and the momentum operator $\hat p$. The set
\begin{equation}
\left\{ \left. \, {\rm e}^{{\rm i}t} \mathcal{D}(\alpha , \beta)\, \,
\right|\, t,\alpha ,\beta \!\in \!\mathbb{R}\, \right\} \qquad \mbox{where}\quad 
\mathcal{D}(\alpha ,\beta ) ={\rm e}^{-\frac{\rm i}{2}\alpha \beta }\,{\rm e}^{i\beta \hat q}\, {\rm e}^{-i\alpha \hat p}
\end{equation}
considered with the multiplication law defined by
\begin{equation}\label{multlaw}
\mathcal{D}(\alpha _1,\beta _1)\, \mathcal{D}(\alpha _2,\beta _2)={\rm e}^{-\frac{\rm i}{2}(\alpha _1 \beta _2-\alpha _2\beta _1)}\, \mathcal{D}(\alpha _1\!+\!\alpha _2,\beta _1\!+\!\beta _2).
\end{equation}
is a group, called the Heisenberg-Weyl group. In the coordinate representation  \cite{Pe}
\begin{equation}
\hat q\psi (q)=q\, \psi (q),
\end{equation}
\begin{equation}
\begin{array}{l}
\hat p =-{\rm i}\frac{d}{dq}
\end{array}
\end{equation}
\begin{equation}\label{Dpsi}
\begin{array}{l}
\mathcal{D}(\alpha ,\beta )\psi (q)={\rm e}^{-\frac{\rm i}{2}\alpha \beta }\,{\rm e}^{i\beta q}\, \psi (q\!-\!\alpha ).
\end{array}
\end{equation}
The operator $\hat p$ satisfies the relation $\hat p =\mathcal{F}^+\hat q\mathcal{F}$ leading to the integral representation
\begin{equation}
 (\hat p \psi )(q)=\frac{1}{2\pi }\int_{\mathbb{R}^2}x\, {\rm e}^{{\rm i}xy}\, \psi (q-y) dx\, dy.
\end{equation}

The linear operator ${\bf Q}:l^2(\mathcal{R}_d)\longrightarrow l^2(\mathcal{R}_d)$ defined by the relation
\begin{equation}
{\bf Q}\varphi (u )=u\,  \varphi (u)\qquad \mbox{for}\qquad u \!\in \!\mathcal{R}_d
\end{equation}
can be regarded as a finite counterpart of $\hat q$. The set $\{ \varepsilon _n\}_{n=-s}^s\subset l^2(\mathcal{R}_d)$, where
\begin{equation}
\varepsilon _n(k\sqrt{\delta})\!=\!\delta _{nk}\qquad \mbox{for}\qquad k\!\in \!\{ -s,-s\!+\!1,...,s\!-\!1,s\}
\end{equation}
is an orthonormal basis, and by using   Dirac's notation, we have
\begin{equation}
\begin{array}{l}
{\bf Q}\!=\!\sum\limits_{n=-s}^sn\sqrt{\delta}\, |\varepsilon_n\rangle  \langle \varepsilon_n| \qquad \mbox{and}\qquad 
{\bf F}\!=\!\frac{1}{\sqrt{d}}\sum\limits_{n,m=-s}^s{\rm e}^{-\frac{2\pi {\rm i}}{d}nm}\, |\varepsilon_n\rangle  \langle \varepsilon_m| .
\end{array}
\end{equation}
The conjugate momentum defined as \cite{MR,Sa,Sc,ST,Vo,Zh}
\begin{equation}
{\bf P}:l^2(\mathcal{R}_d)\longrightarrow l^2(\mathcal{R}_d),\qquad {\bf P} ={\bf F}^+{\bf Q}{\bf F}
\end{equation}
satisfies the relations
\begin{equation}
\begin{array}{l}
({\bf P}\varphi )(u)\!=\!\frac{1}{d}\sum\limits_{\alpha ,v\in \mathcal{R}_d}\alpha {\rm e}^{{\rm i}\alpha  v}\varphi (u\!-\!v)\\[2mm]
 ({\rm e}^{-i\alpha \,{\bf P}}\varphi )(u)\!=\!\varphi (u\!-\!\alpha )
\end{array}
\end{equation}
and 
\begin{equation}
{\rm e}^{-i\sqrt{\delta } \,{\bf P}}|\varepsilon _n\rangle =|\varepsilon _{n+1}\rangle .
\end{equation}
The {\em finite phase space} $\mathcal{R}_d^2$ is a discrete counterpart of $\mathbb{R}^2$ and the unitary operators
\begin{equation}
\begin{array}{l}
{\bf D}(\alpha ,\beta ) 
={\rm e}^{-\frac{\rm i}{2}\alpha \beta }\,{\rm e}^{i\beta \, {\bf Q}}\, {\rm e}^{-i\alpha \,{\bf P}}\qquad \mbox{where}\qquad (\alpha ,\beta )\!\in \! \mathcal{R}_d^2
\end{array}
\end{equation}
represent a finite counterpart for $\mathcal{D}(\alpha ,\beta ) $. They satisfy the relations \cite{Vo}
\begin{equation}
\begin{array}{l}
{\bf D}(\alpha ,\beta )\varphi  (u)={\rm e}^{-\frac{\rm i}{2}\alpha \beta }\,{\rm e}^{i\beta u}\, \varphi (u\!-\!\alpha )
\end{array}
\end{equation}
\begin{equation}
{\bf D}(\alpha _1,\beta _1)\, {\bf D}(\alpha _2,\beta _2) 
={\rm e}^{-\frac{\rm i}{2}(\alpha _1 \beta _2-\alpha _2\beta _1)}{\bf D}(\alpha _1\!+\!\alpha _2,\beta _1\!+\!\beta _2) 
\end{equation}
and define a projective representation of  a finite version of the Heisenberg-Weyl group.

\section{Ground state of the harmonic oscillator and a finite counterpart}

Let $\kappa \!\in \!(0,\infty )$. It is well-known that the Gaussian function 
\begin{equation}\label{cgauss}
g_\kappa :\mathbb{R}\longrightarrow \mathbb{R}, \qquad g_\kappa (x)
={\rm e}^{-\frac{\kappa }{2}x^2}
\end{equation}
satisfies the relation
\begin{equation}\label{Fcont}
 \frac{1}{\sqrt{2\pi }}\int_{-\infty }^{\infty}\mathrm{e}^{-\mathrm{i}\xi x} 
\mathrm{e}^{-\frac{\kappa }{2}x^2}\,  dx=\frac{1}{\sqrt{\kappa }}\, \mathrm{e}^{-\frac{1}{2\kappa }\xi ^2}
\end{equation}
that is,
\begin{equation}\label{Fgaussian}
\begin{array}{l}
\mathcal{F}[g_\kappa  ]=\frac{1}{\sqrt{\kappa }}\,g_{\frac{1}{\kappa }}.
\end{array}
\end{equation}
The norm of $g_1$ is $|| g_1||=\sqrt[4]{\pi }$, and the normalized function $g_1/||g_1||$, namely,
\begin{equation}
\Psi _0:\mathbb{R}\longrightarrow \mathbb{R}, \qquad \Psi _0(x)=\frac{1}{\sqrt[4]{\pi }}{\rm e}^{-\frac{1}{2}x^2}
\end{equation}
is the ground state of the harmonic oscillator in the coordinate representation.\\

\noindent
{\bf Lemma 1}. {\it We have}
\begin{equation}
\sum_{\ell =-\infty }^\infty \mathrm{e}^{-\frac{\kappa \pi }{d} \left(\ell d +x\right)^2}
=\frac{1}{\sqrt{\kappa d}}\,\sum_{\ell =-\infty }^\infty   \mathrm{e}^{\frac{2\pi \mathrm{i}}{d}\ell x}\, 
\mathrm{e}^{-\frac{\pi}{\kappa d}\ell ^2}\qquad \mbox{\it for any}\quad x\!\in \!\mathbb{R}. 
\end{equation}
{\bf Proof}. The periodic function
\[
G_\kappa :\mathbb{R}\longrightarrow \mathbb{R},\qquad G_\kappa (x)
=\sum_{\alpha =-\infty }^\infty \mathrm{e}^{-\frac{\kappa \pi }{d} \left(\alpha d +x\right)^2}
\]
with period $d$ admits the Fourier expansion
\[
G_\kappa (x)=\sum_{\ell =-\infty }^\infty a_\ell \,  \mathrm{e}^{\frac{2\pi \mathrm{i}}{d}\ell x}
\]
where
\[ \fl
a_\ell =\frac{1}{d}\int _0^d\mathrm{e}^{-\frac{2\pi \mathrm{i}}{d}\ell x}\sum_{\alpha =-\infty }^\infty \mathrm{e}^{-\frac{\kappa }{2}\left(\sqrt{\frac{2\pi }{d}}\, (\alpha d +x)\right)^2}\, dx=\frac{1}{d}\sum_{\alpha =-\infty }^\infty\int _0^d\mathrm{e}^{-\frac{2\pi \mathrm{i}}{d}\ell x} \mathrm{e}^{-\frac{\kappa }{2}\left(\sqrt{\frac{2\pi }{d}}\, (\alpha d +x)\right)^2}\, dx
\]
By denoting $t\!=\!\sqrt{\frac{2\pi }{d}}\, (\alpha d \!+\!x)$ and using (\ref{Fcont}) we get \cite{Me}
\[
\begin{array}{rl}
a_\ell & =\frac{1}{\sqrt{2\pi d}}\sum_{\alpha =-\infty }^\infty \int_{\alpha \sqrt{2\pi d}}^{(\alpha +1)\sqrt{2\pi d}}\mathrm{e}^{-\frac{2\pi \mathrm{i}}{d}\ell \left(t\sqrt{\frac{d}{2\pi }}-\alpha d\right)}
\mathrm{e}^{-\frac{\kappa }{2}t^2}\, dt\\[4mm]
 & =\frac{1}{\sqrt{2\pi d}}\sum_{\alpha =-\infty }^\infty \int_{\alpha \sqrt{2\pi d}}^{(\alpha +1)\sqrt{2\pi d}}\mathrm{e}^{-\mathrm{i}\ell  t\sqrt{\frac{2\pi }{d}}}
\mathrm{e}^{-\frac{\kappa }{2}t^2}\, dt\\[4mm]
& =\frac{1}{\sqrt{2\pi d}}\int_{-\infty }^{\infty}\mathrm{e}^{-\mathrm{i}\ell  t\sqrt{\frac{2\pi }{d}}} 
\mathrm{e}^{-\frac{\kappa }{2}t^2}\,  dt=\frac{1}{\sqrt{\kappa d}}\, \mathrm{e}^{-\frac{\pi}{\kappa d}\ell ^2}
\end{array}
\]
whence
\[
G_\kappa (x)=\frac{1}{\sqrt{\kappa d}}\,\sum_{\ell =-\infty }^\infty   \mathrm{e}^{\frac{2\pi \mathrm{i}}{d}\ell x}\, 
\mathrm{e}^{-\frac{\pi}{\kappa d}\ell ^2}.\qquad \opensquare 
\]

The periodic function $\mathfrak{g}_\kappa  :\mathbb{Z}\sqrt{\delta}\longrightarrow \mathbb{R}$,
\begin{equation}
\mathfrak{g}_\kappa  (n\sqrt{\delta})=\sum_{\ell =-\infty }^\infty {\rm e}^{-\frac{\kappa \pi }{d}(\ell d+n)^2}
=\frac{1}{\sqrt{\kappa d}}\,\sum_{\ell =-\infty }^\infty   \mathrm{e}^{\frac{2\pi \mathrm{i}}{d}\ell n}\, 
\mathrm{e}^{-\frac{\pi}{\kappa d}\ell ^2}
\end{equation}
defined by using a Zak type transformation \cite{Zak}, satisfies the relation \cite{CD}
\[
\begin{array}{rl}
\mathfrak{g}_\kappa (j\sqrt{\delta}) & =G_\kappa (j)=\frac{1}{\sqrt{\kappa d}}\,\sum_{\ell =-\infty }^\infty   \mathrm{e}^{\frac{2\pi \mathrm{i}}{d}j\ell }\, \mathrm{e}^{-\frac{\pi}{\kappa d}\ell ^2}\\[3mm]
 & =\frac{1}{\sqrt{\kappa d}}\sum_{n=-s}^{s}\sum_{\alpha =-\infty }^\infty  \mathrm{e}^{\frac{2\pi \mathrm{i}}{d}j(\alpha d+n) }\, \mathrm{e}^{-\frac{\pi }{\kappa d}(\alpha d+n) ^2}\, \\[2mm]
& =\frac{1}{\sqrt{\kappa }}\frac{1}{\sqrt{d}}\sum_{n=-s}^{s}\mathrm{e}^{\frac{2\pi \mathrm{i}}{d}jn }\sum_{\alpha =-\infty }^\infty  \, \mathrm{e}^{-\frac{\pi }{\kappa d}(\alpha d+n) ^2}
\end{array}
\]
equivalent with
\begin{equation}\label{Fourier-Gauss}
\begin{array}{l}
\mathbf{F}[\mathfrak{g}_\kappa  ]=\frac{1}{\sqrt{\kappa }}\,\mathfrak{g}_{\frac{1}{\kappa }}.
\end{array}
\end{equation}
This equality has been obtained by Ruzzi \cite{Ru} by using the relation
\begin{equation}
\begin{array}{l}
\mathfrak{g}_\kappa  (n\sqrt{\delta})=\frac{1}{\sqrt{\kappa d}}\, \theta _3\left( \frac{n}{d},\frac{{\rm i}}{\kappa d} \right)
\end{array}
\end{equation}
and the properties of the Jacobi theta function
\begin{equation}
\theta _3(z,\tau )=\sum_{\alpha =-\infty }^\infty {\rm e}^{{\rm i}\pi \tau \alpha ^2}\, {\rm e}^{2\pi {\rm i}\alpha z}.
\end{equation}
\mbox{}\\[3mm]
\noindent
{\bf Lemma 2}. {\it If the numbers $N_{r,t}$ are such that the series are absolutely  convergent then} 
\begin{equation}
\sum\limits_{r,t =-\infty }^\infty N_{r,t}=\sum\limits_{k  ,\ell =-\infty }^\infty N_{k +\ell ,k -\ell } + \sum\limits_{k  ,\ell  =-\infty }^\infty N_{k+\ell +1,k -\ell }.
\end{equation}
{\bf Proof}. We separate the sum as \cite{Marzoli}
\[
\sum\limits_{r,t =-\infty }^\infty N_{r,t }=\sum\limits_{\scriptsize 
\begin{array}{c}
r,t \\
{\rm both\ even}\\
{\rm or}\\
{\rm both\ odd}
\end{array}} N_{r,t }+\sum\limits_{\scriptsize 
\begin{array}{c}
r,t \\
{\rm one\ even}\\
{\rm and}\\
{\rm other\ odd}
\end{array}} N_{r,t }
\]
and use the substitutions $(r,t )\!=\!(k \!+\!\ell ,k \!-\!\ell )$ and $(r,t)\!=\!(k \!+\!\ell \!+\!1,k \!-\!\ell )$.\qquad \opensquare
\mbox{}\\[3mm]

The function $\mathfrak{g}_1$ is a finite counterpart of $g_1$ and $g_1 ^2=g_{2 }$, but $\mathfrak{g}_1 ^2\not=\mathfrak{g}_{2 }$.\\[5mm]
\noindent
{\bf Theorem 1}. {\it We have}
\begin{equation}
\mathfrak{g}_1^2(n\sqrt{\delta })= \left(2\, \mathfrak{g}_2(0)\!-\!\mathfrak{g}_{\frac{1}{2}}(0)\right) \mathfrak{g}_2(n\sqrt{\delta })\!-\!\left(\mathfrak{g}_2(0) \!-\!\mathfrak{g}_{\frac{1}{2}}(0)\right)\mathfrak{g}_{\frac{1}{2}}(2n\sqrt{\delta }).
\end{equation}
{\bf Proof}. By using Lemma 2 we get 
\[
\begin{array}{rl}
\mathfrak{g}_1^2(n\sqrt{\delta }) & \!\!\!\!= \frac{1}{d}\,\sum\limits_{r,t =-\infty }^\infty   \mathrm{e}^{\frac{2\pi \mathrm{i}}{d}(r+t) n}\, 
\mathrm{e}^{-\frac{\pi}{d}r^2}\, \mathrm{e}^{-\frac{\pi}{d}t ^2}\\[5mm]
& \!\!\!\!= \frac{1}{d}\,\sum\limits_{k =-\infty }^\infty   \mathrm{e}^{\frac{2\pi \mathrm{i}}{d}k \, 2n}\, 
\mathrm{e}^{-\frac{2\pi}{d}k^2}\,\sum\limits_{\ell =-\infty }^\infty  \mathrm{e}^{-\frac{2\pi}{d}\ell ^2}\\[5mm]
& \!\!+ \frac{1}{d}\,\sum\limits_{k =-\infty }^\infty   \mathrm{e}^{\frac{2\pi \mathrm{i}}{d}(2k+1)n}\, 
\mathrm{e}^{-\frac{\pi}{2d}(2k+1)^2}\,\sum\limits_{\ell =-\infty }^\infty  \mathrm{e}^{-\frac{\pi}{2d}(2\ell +1)^2}\\[5mm]
& \!\!\!\!= \left(2\, \mathfrak{g}_2(0)\!-\!\mathfrak{g}_{\frac{1}{2}}(0)\right) \mathfrak{g}_2(n\sqrt{\delta })\!-\!\left(\mathfrak{g}_2(0) \!-\!\mathfrak{g}_{\frac{1}{2}}(0)\right)\mathfrak{g}_{\frac{1}{2}}(2n\sqrt{\delta }).\qquad \opensquare 
\end{array}
\]
With the exception of a few small values of $d$, we have $\mathfrak{g}_2(0)\approx 1\approx \mathfrak{g}_{\frac{1}{2}}(0)$, and hence 
$\mathfrak{g}_1^2\approx \mathfrak{g}_2$. For example in the case $d\!=\!21$, we have $\mathfrak{g}_2(0)\approx 1+10^{-57}$ and $\mathfrak{g}_{\frac{1}{2}}(0)\approx  1+10^{-14}$.

The norm  $\mathcal{N}=||\mathfrak{g}_1||$ of $\mathfrak{g}_1$ satisfies the relation
\begin{equation}\label{normofg}
\fl
\begin{array}{rl}
 \mathcal{N}^2 & \!\!\!\!=\!\frac{1}{d}\sum\limits_{n=-s}^s\left(\sum\limits_{r =-\infty }^\infty  
 \mathrm{e}^{\frac{2\pi \mathrm{i}}{d}r n}\, 
\mathrm{e}^{-\frac{\pi}{d}r^2}\right)^2\!\!\!=\!\frac{1}{d}\sum\limits_{n=-s}^s\, \, \sum\limits_{r,t=-\infty }^\infty \! 
\mathrm{e}^{\frac{2\pi \mathrm{i}}{d}(r+t)n}\, \mathrm{e}^{-\frac{\pi}{d}r^2}\, \mathrm{e}^{-\frac{\pi}{d}t^2}\\[4mm]
& \!\!\!\!=\!\frac{1}{d}\sum\limits_{r,t=-\infty }^\infty \, \sum\limits_{n=-s}^s
\mathrm{e}^{\frac{2\pi \mathrm{i}}{d}(r+t)n}\, \mathrm{e}^{-\frac{\pi}{d}r^2}\, \mathrm{e}^{-\frac{\pi}{d}t^2}=
\sum\limits_{r=-\infty }^\infty \!\mathrm{e}^{-\frac{\pi}{d}r^2} 
\sum\limits_{\ell =-\infty }^\infty \! \mathrm{e}^{-\frac{\pi}{d}(\ell d-r) ^2}.
\end{array}
\end{equation}
The finite counterpart of $\Psi _0$ is the periodic function  $(1/\mathcal{N})\mathfrak{g}_1$, that is, the function
\begin{equation}
\begin{array}{rl}
{\bf g}  :\mathbb{Z}\sqrt{\delta}\longrightarrow \mathbb{R},\qquad 
 {\bf g}(n\sqrt{\delta})
 & \!\!\!\!=\frac{1}{\mathcal{N}}\,\sum\limits_{\ell =-\infty }^\infty {\rm e}^{-\frac{\pi }{d}(\ell d+n)^2}\\[3mm]
& \!\!\!\!=\frac{1}{\mathcal{N}\sqrt{d}}\,\sum\limits_{\ell =-\infty }^\infty   \mathrm{e}^{\frac{2\pi \mathrm{i}}{d}\ell n}\, 
\mathrm{e}^{-\frac{\pi}{d}\ell ^2}.
 \end{array}
\end{equation}

\noindent
{\bf Theorem 2}. {\it For any $n,m\!\in \!\{ -s,-s\!+\!1,...,s\!-\!1,s\}$ we have}
\begin{equation}
\begin{array}{l}
\frac{1}{\sqrt{d}}\, \sum\limits_{a=-s}^s{\bf g}((n-a)\sqrt{\delta})\, {\bf g}((m-a)\sqrt{\delta})= {\bf F}[{\bf g}^2]((n\!-\!m)\sqrt{\delta }).
\end{array}
\end{equation}
{\bf Proof}.  By using the definition of ${\bf g}$ we get
\[\fl
\begin{array}{l}
\sum\limits_{a=-s}^s{\bf g}((n-a)\sqrt{\delta})\, {\bf g}((m-a)\sqrt{\delta})\!= \!
\frac{1}{d\mathcal{N}^2}\sum\limits_{r,t =-\infty }^\infty 
\sum\limits_{a=-s}^s {\rm e}^{-\frac{2\pi {\rm i}}{d}a (r+t)}
\,{\rm e}^{\frac{2\pi {\rm i}}{d}(r n+tm)} {\rm e}^{-\frac{\pi }{d}r^2}\, {\rm e}^{-\frac{\pi }{d}t^2}\\[4mm]
\qquad \qquad \qquad = \!
\frac{1}{\mathcal{N}^2} \sum\limits_{r=-\infty }^\infty 
\!{\rm e}^{\frac{2\pi {\rm i}}{d}r(n-m)} {\rm e}^{-\frac{\pi }{d}r^2}
\sum\limits_{\ell =-\infty }^\infty {\rm e}^{-\frac{\pi }{d}(\ell d-r)^2}\\[4mm]
\qquad \qquad \qquad = \!
\frac{1}{\mathcal{N}^2}\sum\limits_{k=-s}^s\sum\limits_{t=-\infty }^\infty 
\!{\rm e}^{\frac{2\pi {\rm i}}{d}(td-k)(n-m)} {\rm e}^{-\frac{\pi }{d}(td-k)^2}
\sum\limits_{\ell =-\infty }^\infty {\rm e}^{-\frac{\pi }{d}(\ell d-td-k)^2}\\[4mm]
\qquad \qquad \qquad = \!
\frac{1}{\mathcal{N}^2} 
\sum\limits_{k=-s}^s{\rm e}^{-\frac{2\pi {\rm i}}{d}k(n-m)}\sum\limits_{t=-\infty }^\infty 
\! {\rm e}^{-\frac{\pi }{d}(td-k)^2}
\sum\limits_{\ell =-\infty }^\infty {\rm e}^{-\frac{\pi }{d}(\ell d-k)^2}\\[4mm]
\qquad \qquad \qquad = \!
\sqrt{d}\, \, {\bf F}[{\bf g}^2]((n\!-\!m)\sqrt{\delta }).\qquad  \opensquare 
\end{array} 
\]

%

\section{Standard coherent states and a finite counterpart}

\begin{figure}[t]
\centering
\includegraphics[scale=0.7]{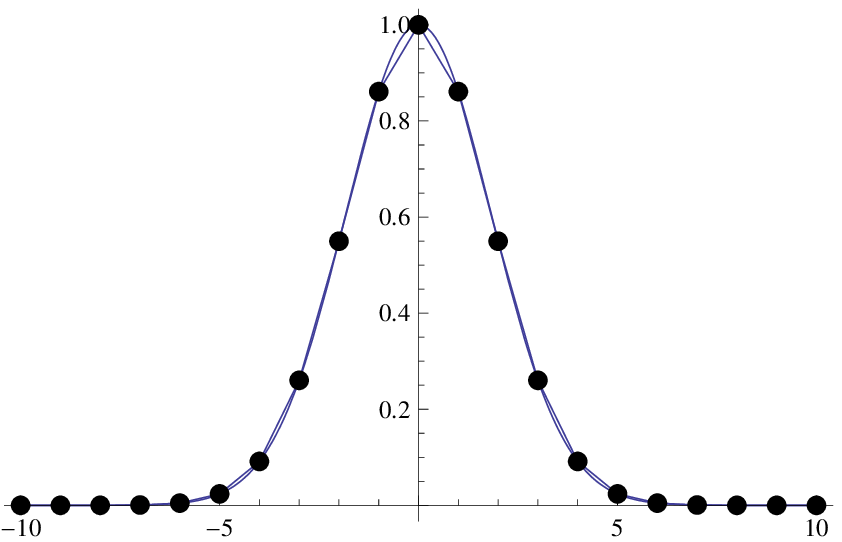}\qquad 
\includegraphics[scale=0.7]{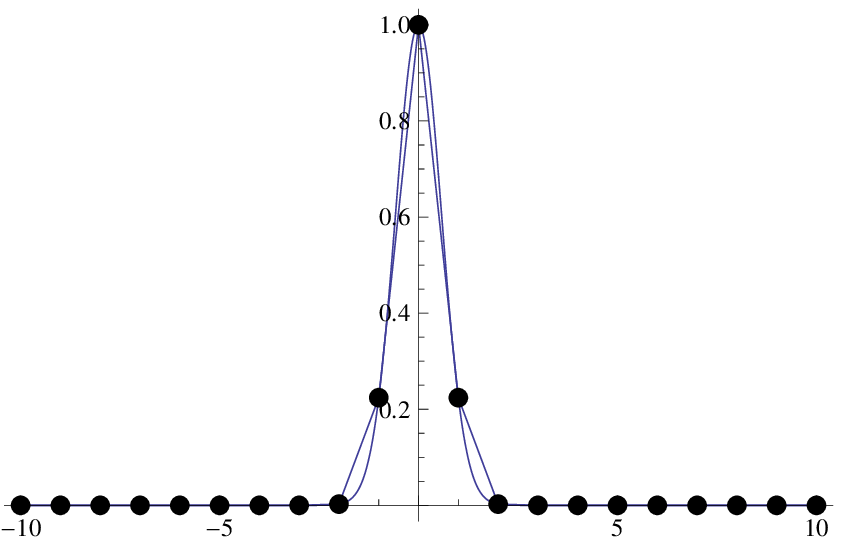}
\caption{\label{gaussians} The functions $g_1$, $\mathfrak{g}_1 $ (left) and $g_{10}$, $\mathfrak{g}_{10} $ (right)
in the case $d\!=\!21$. }
\end{figure}

The standard coherent states can be defined as the states
\begin{equation}
 |\alpha ,\beta \rangle \!=\!\mathcal{D}(\alpha ,\beta )\Psi _0 \qquad \mbox{with}\qquad (\alpha ,\beta )\in \mathbb{R}^2.
\end{equation}
They satisfy in $L^2(\mathbb{R})$ the resolution of the identity 
\begin{equation}
 \mathbb{I}=\frac{1}{2\pi }\int_{\mathbb{R}^2}d\alpha \, d\beta \, |\alpha ,\beta \rangle \langle \alpha ,\beta |
\end{equation}
that is, we have
\[
 |\psi \rangle =\frac{1}{2\pi }\int_{\mathbb{R}^2}d\alpha \, d\beta \, |\alpha ,\beta \rangle \langle \alpha ,\beta |\psi \rangle \qquad \mbox{for any}\qquad \psi \in L^2(\mathbb{R}).
\]
For $d\!=\!2s\!+\!1$ large enough, the norm $\mathcal{N}$ of $\mathfrak{g}_1$ satisfies the relation 
\begin{equation}
\begin{array}{l}
 \mathcal{N}=\sqrt{\sum_{u\in \mathcal{R}_d}(\mathfrak{g}_1(u))^2}\approx \sqrt[4]{\frac{d}{2}}.
\end{array} 
\end{equation}
For example, in the case $d\!=\!21$ we have (see Figure \ref{gaussians})
\[
\begin{array}{l}
\max\limits_{u \in \mathcal{R}_d}
\left |\mathfrak{g}_1 (u )\!-\!g_1(u)\right|=1.3\cdot 10^{-8}\qquad \mbox{and}\qquad 
\mathcal{N}\!-\!\sqrt[4]{\frac{d}{2}}=1.7\cdot 10^{-14}.
\end{array}
\]

For $d$ large enough we have
\begin{equation}
\begin{array}{l}
{\bf g}(u)\approx \sqrt[4]{\delta }\, \Psi _0(u)\qquad \mbox{for}\qquad u\!\in \!\mathcal{R}_d.
\end{array}
\end{equation}
The $d^2$ states $\{ |\alpha ,\beta \rangle _{\!\mbox{}_d}\}_{(\alpha ,\beta )\in \mathcal{R}_d^2}$,  where
\begin{equation}
|\alpha ,\beta \rangle _{\!\mbox{}_d} \!=\!{\bf D}(\alpha ,\beta ){\bf g}\qquad \mbox{with}\qquad 
{\bf D}(\alpha ,\beta ){\bf g}(u)\!=\!{\rm e}^{-\frac{\rm i}{2}\alpha \beta }{\rm e}^{{\rm i}\beta u}{\bf g}(u\!-\!\alpha )
\end{equation}
that is, the states
\begin{equation}
\begin{array}{l}
|\alpha ,\beta \rangle _{\!\mbox{}_d}
={\rm e}^{-\frac{\rm i}{2}\alpha \beta }\sum\limits_{n=-s}^s{\rm e}^{{\rm i}\sqrt{\delta }\beta n}{\bf g}(n\sqrt{\delta }-\alpha )\, |\varepsilon _n\rangle 
\end{array}
\end{equation}
satisfy the resolution of the identity in $l^2(\mathcal{R}_d)$ \cite{CGV,TC,Vo}
\begin{equation}
{\bf I}=\frac{1}{d}\sum_{(\alpha ,\beta )\in \mathcal{R}_d^2}|\alpha ,\beta \rangle _{\!\mbox{}_d}\, {}_{\mbox{}_d} \!\langle \alpha ,\beta |. 
\end{equation}
For any $\varphi \!\in \!l^2(\mathcal{R}_d)$ we have
\begin{equation}
\begin{array}{l}
|\varphi \rangle ={\bf I}|\varphi \rangle =\frac{1}{d}\sum\limits_{(\alpha ,\beta )\in \mathcal{R}_d^2}|\alpha ,\beta \rangle _{\!\mbox{}_d}\, {}_{\mbox{}_d} \!\langle \alpha ,\beta |\varphi \rangle 
\end{array}
\end{equation}
and
\begin{equation}
\begin{array}{l}
||\varphi ||^2 =\frac{1}{d}\sum\limits_{(\alpha ,\beta )\in \mathcal{R}_d^2}| {}_{\mbox{}_d} \!\langle \alpha ,\beta |\varphi \rangle |^2.
\end{array}
\end{equation}
The elements of the frame  \cite{Ch,HL} \, $\{|\alpha ,\beta \rangle _{\!\mbox{}_d}\}_{(\alpha ,\beta )\in \mathcal{R}_d^2}$,
in general, are not orthogonal
\begin{equation} 
\begin{array}{l}
{}_{\mbox{}_d} \!\langle \alpha _1,\beta _1 |\alpha _2,\beta _2\rangle _{\!\mbox{}_d}
\!=\!{\rm e}^{\frac{\rm i}{2}(\alpha _1\beta _1-\alpha _2\beta _2)}
\sum\limits_{u\in \mathcal{R}_d}{\rm e}^{{\rm i}(\beta _2-\beta _1)u}\, {\bf g}(u\!-\!\alpha _1)\, 
{\bf g}(u\!-\!\alpha _2).
\end{array}
\end{equation}

\begin{figure}[t]
\centering
\includegraphics[scale=0.7]{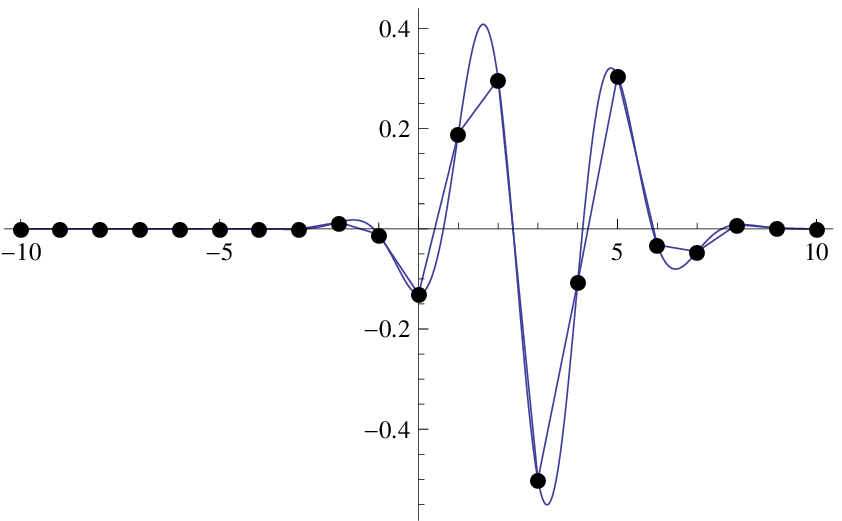}\qquad 
\includegraphics[scale=0.7]{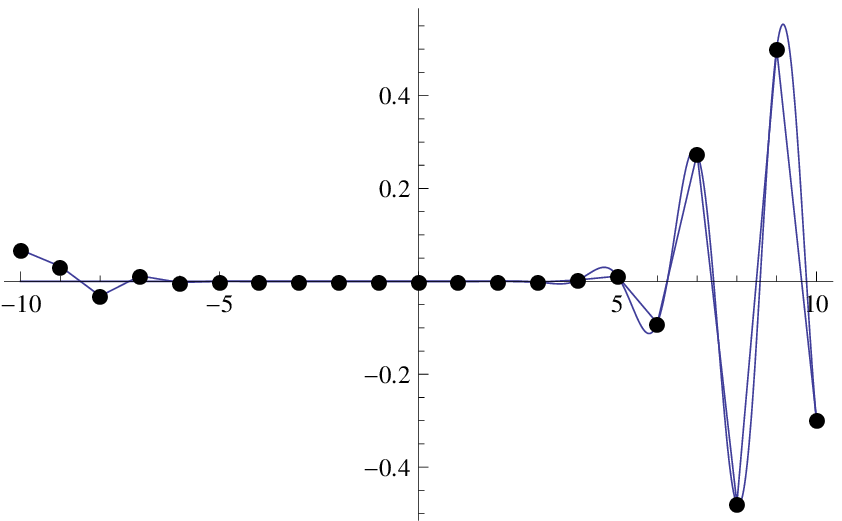}\\[2mm]
\includegraphics[scale=0.7]{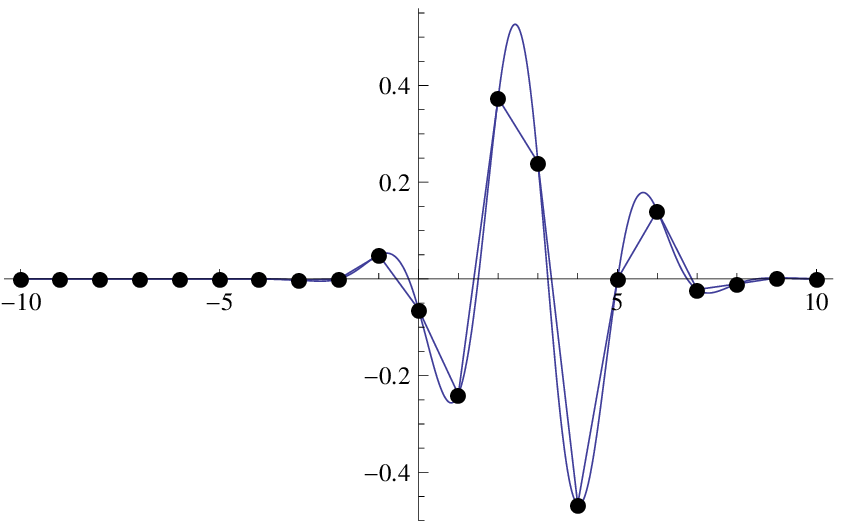}\qquad 
\includegraphics[scale=0.7]{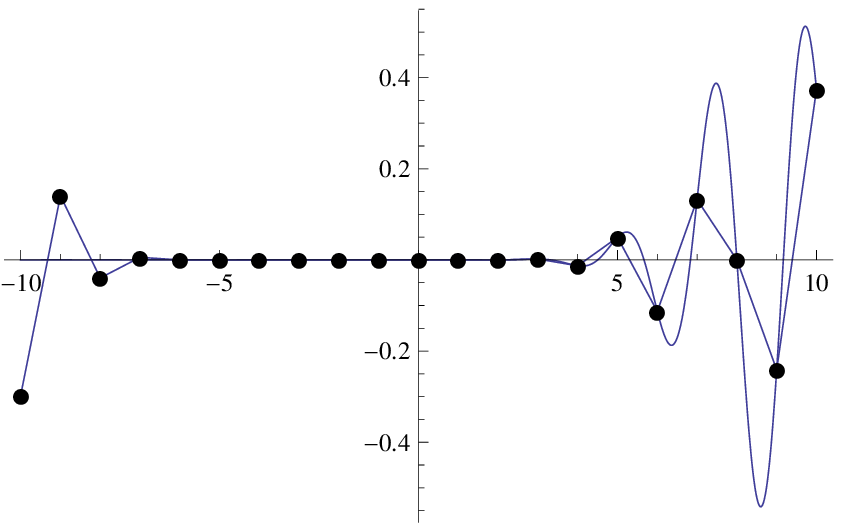}
\caption{\label{fig2} The real part (first line) and the imaginary part (second line) of 
$|3\sqrt{\delta } ,6\sqrt{\delta } \rangle $ versus $|3\sqrt{\delta } ,6\sqrt{\delta } \rangle _{\!\mbox{}_{21}}$ 
(left) and $|9\sqrt{\delta } ,9\sqrt{\delta } \rangle $ versus 
$|9\sqrt{\delta } ,9\sqrt{\delta } \rangle _{\!\mbox{}_{21}}$ (right).}
\end{figure}

The finite frame $\{|\alpha ,\beta \rangle _{\!\mbox{}_d}\}_{(\alpha ,\beta )\in \mathcal{R}_d^2}$ is a 
finite counterpart of the system of standard coherent states 
$\{ |\alpha ,\beta \rangle \}_{(\alpha ,\beta )\in \mathbb{R}^2}$, 
and for $d$ large enough we have
\begin{equation}\label{approxcs1}
\begin{array}{l}
{\bf D}(\alpha ,\beta ){\bf g}(u)\approx \sqrt[4]{\delta }\, \mathcal{D}(\alpha ,\beta )\Psi _0(u)\qquad \mbox{for almost all}\quad u\!\in \!\mathcal{R}_d.
\end{array}
\end{equation}
The agreement is very good in the midle part of $\mathcal{R}_d$, 
but some significant differences may occur in the extreme parts  (see Figure \ref{fig2}). 
In the particular case $d\!=\!21$ we have $\mathcal{R}_{21}\!=\!\{ n\sqrt{\delta }\ \ |\ -10\leq n\leq 10 \}$ with $\delta =2\pi /21$.
The values of 
\[
\max _{-8\leq n\leq 8} \left|{\bf D}(\alpha ,\beta ){\bf g}(n\sqrt{\delta })
- \sqrt[4]{\delta }\, \mathcal{D}(\alpha ,\beta )\Psi _0(n\sqrt{\delta })\right|
\]
in certain particular cases are presented in Table 1 . In the case $(\alpha ,\beta )\in \mathcal{R}_d^2$, 
we consider that the restriction of the function $\sqrt[4]{\delta }\, |\alpha ,\beta \rangle $ to $\mathcal{R}_d$ 
is almost identical to $|\alpha ,\beta \rangle _{\!\mbox{}_d}$. As concern the Fourier transform
\begin{equation}\label{Fouriercs}
\mathcal{F}|\alpha ,\beta \rangle =|-\beta , \alpha \rangle \qquad \mbox{and}\qquad 
{\bf F}|\alpha ,\beta \rangle _{\!\mbox{}_d}=|-\beta , \alpha \rangle _{\!\mbox{}_d}.
\end{equation}
Since ${\bf F}$ is a unitary operator, we have 
\begin{equation}
{}_{\mbox{}_d} \!\langle \alpha _1,\beta _1 |\alpha _2,\beta _2\rangle _{\!\mbox{}_d}={}_{\mbox{}_d} \!\langle -\beta_1,\alpha _1 |-\beta _2,\alpha _2 \rangle _{\!\mbox{}_d}.
\end{equation}

\begin{table}
\caption{ Differences between continuous and discrete coherent states.}
\begin{indented}
\lineup
\item[]\begin{tabular}{ccccc}
\br
  & $\beta \!=\!\sqrt{\delta }$  & $\beta \!=\!3\sqrt{\delta}$ & $\beta \!=\!6\sqrt{\delta}$ & $\beta \!=\!9\sqrt{\delta}$ \\
\mr
$\alpha \!=\!\sqrt{\delta}$ &  $2.44895\cdot 10^{-10}$& $2.44895\cdot 10^{-10}$ & $2.44894\cdot 10^{-10}$& $2.43229\cdot 10^{-9}$ \\
$\alpha \!=\!3\sqrt{\delta}$ &  $1.76877\cdot 10^{-7}$& $1.76877\cdot 10^{-7}$ & $1.76877\cdot 10^{-7}$& $1.70238\cdot 10^{-7}$ \\
$\alpha \!=\!6\sqrt{\delta}$ &  $0.000364047$& $0.000364047$ & $0.000364047$& $0.000364667$ \\
$\alpha \!=\!9\sqrt{\delta}$ &  $0.0507198$& $0.0507198$ & $0.0507199$& $0.0507748$ \\
\br
\end{tabular}
\end{indented}
\end{table}

\section{Finite oscillator obtained by using finite difference operators}

The Hamiltonian of the harmonic oscillator
\begin{equation}
\hat H=-\frac{1}{2}\, \frac{{\rm d}^2}{{\rm d}x^2}+\frac{1}{2}x^2
=-\frac{1}{2}(\mathcal{D}^2+\mathcal{F}\mathcal{D}^2\mathcal{F}^+).
\end{equation}
satisfies the relation
\begin{equation}
 \mathcal{F}\hat H=\hat H\mathcal{F}
\end{equation}
and the system $\{\Psi _n\}_{n\in \mathbb{N}}$ of Hermite-Gaussian functions  
\begin{equation}
\begin{array}{l}
\Psi _n:\mathbb{R}\longrightarrow \mathbb{R},\qquad 
\Psi _n(x)=\frac{1}{\sqrt{n!\, 2^n\sqrt{\pi }}}\, H_n(x)\, {\rm e}^{-\frac{1}{2}x^2}.
\end{array}
\end{equation}
denoted sometimes by $\{ |n\rangle \}_{n\in \mathbb{N}}$, is a complete othonormal  set of common eigenfunctions
\begin{equation}
\begin{array}{ll}
\langle n|k\rangle =\delta _{nk}, & \hat H|n\rangle\!=\!\left(n\!+\!\frac{1}{2}\right)|n\rangle,\\[2mm]
\sum\limits_{n=0}^\infty |n\rangle \langle n|\!=\!\mathbb{I},\qquad  & \mathcal{F}|n\rangle =(-{\rm i})^n|n\rangle .
\end{array}
\end{equation}

The finite-difference operator $\tilde \mathcal{D}^2$, where
\begin{equation}
\tilde \mathcal{D}^2\varphi (x)=\frac{\varphi (x+\epsilon )-2\varphi (x)+\varphi (x-\epsilon )}{\epsilon ^2}
\end{equation}
is an approximation of $\mathcal{D}^2$, and for $x=n\epsilon $ with $\epsilon =\sqrt{\delta}$ we get
\begin{equation}
\begin{array}{l}
{\bf F}\tilde \mathcal{D}^2{\bf F}^+=2d\left(\cos \frac{2\pi n}{d}-1\right).
\end{array}
\end{equation}
The finite-difference Hamiltonian
\begin{equation}
\mathcal{H}_d=\tilde \mathcal{D}^2+{\bf F}\tilde \mathcal{D}^2{\bf F}^+
\end{equation}
commutes with the finite Fourier transform
\begin{equation}
{\bf F}\mathcal{H}_d=\mathcal{H}_d{\bf F}.
\end{equation}
By denoting $\varphi[n]:=\varphi (n\sqrt{\delta })$, the equation $\mathcal{H}_d\varphi =\lambda \varphi $ becomes
\begin{equation}
\begin{array}{l}
\varphi [n\!+\!1]-2\varphi [n]+\varphi [n\!-\!1]+2\left( \cos \frac{2\pi n}{d}-1\right)\varphi [n]=\lambda \varphi [n].
\end{array}
\end{equation}
It is known \cite{Ba} that the eigenvalues of $\mathcal{H}_d$ are distinct for odd $d$ (see Figure \ref{eigval}). Therefore,  
the eigenfunctions of $\mathcal{H}_d$ are at the same time eigenfunctions of ${\bf F}$.
The normalized eigenfunctions ${\bf h}_m$ of $\mathcal{H}_d$, considered in the increasing order of the 
number of sign alternations, satisfy the relation \cite{Ba,WK}
\begin{equation}
 {\bf F}\, {\bf h}_m=(-{\rm i})^m\, {\bf h}_m
\end{equation}
and are called {\em Harper functions}. They correspond to the eigenvectors of the matrix
\begin{equation}
\begin{array}{l}
\left(2 ( \cos \frac{2 \pi n}{d} \!-\! 2) \delta_{nm}  + 
 \delta_{n, m + 1} + \delta_{n, m - 1} + 
 \delta_{n, m - 2  s} + \delta_{n,m + 2  s}
\right)_{-s\leq n,m\leq s}
\end{array}
\end{equation}
and can be regarded as a finite version of Hermite-Gaussian functions $\Psi _0,\Psi _1,...,\Psi _{d-1}$.

\begin{center}
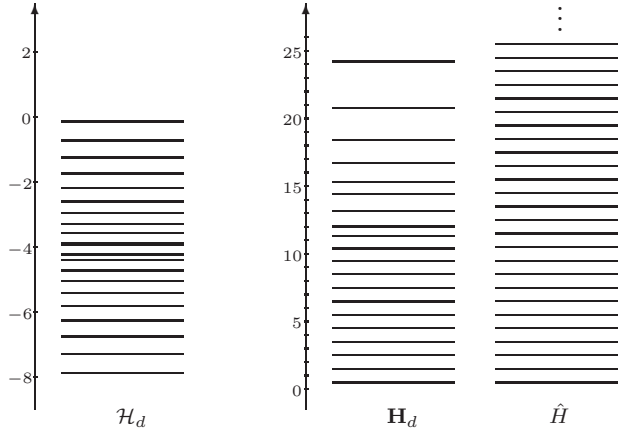
\begin{figure}[t] 
\setlength{\unitlength}{1.8mm}
\begin{picture}(80,32)(10,0)
\put(30,3){\vector(0,1){30}}
\put(50,3){\vector(0,1){30}}

\put(49.9,4.5){\line(1,0){0.3}}
\put(49.9,5.5){\line(1,0){0.3}}
\put(49.9,6.5){\line(1,0){0.3}}
\put(49.9,7.5){\line(1,0){0.3}}
\put(49.9,8.5){\line(1,0){0.3}}
\put(49.9,9.5){\line(1,0){0.3}}
\put(49.9,10.5){\line(1,0){0.3}}
\put(49.9,11.5){\line(1,0){0.3}}
\put(49.9,12.5){\line(1,0){0.3}}
\put(49.9,13.5){\line(1,0){0.3}}
\put(49.9,14.5){\line(1,0){0.3}}
\put(49.9,15.5){\line(1,0){0.3}}
\put(49.9,16.5){\line(1,0){0.3}}
\put(49.9,17.5){\line(1,0){0.3}}
\put(49.9,18.5){\line(1,0){0.3}}
\put(49.9,19.5){\line(1,0){0.3}}
\put(49.9,20.5){\line(1,0){0.3}}
\put(49.9,21.5){\line(1,0){0.3}}
\put(49.9,22.5){\line(1,0){0.3}}
\put(49.9,23.5){\line(1,0){0.3}}
\put(49.9,24.5){\line(1,0){0.3}}
\put(49.9,25.5){\line(1,0){0.3}}
\put(49.9,26.5){\line(1,0){0.3}}
\put(49.9,27.5){\line(1,0){0.3}}
\put(49.9,28.5){\line(1,0){0.3}}
\put(49.9,29.5){\line(1,0){0.3}}
\put(49.9,30.5){\line(1,0){0.3}}

\put(64,5){\line(1,0){9}}
\put(64,6){\line(1,0){9}}
\put(64,7){\line(1,0){9}}
\put(64,8){\line(1,0){9}}
\put(64,9){\line(1,0){9}}
\put(64,10){\line(1,0){9}}
\put(64,11){\line(1,0){9}}
\put(64,12){\line(1,0){9}}
\put(64,13){\line(1,0){9}}
\put(64,14){\line(1,0){9}}
\put(64,15){\line(1,0){9}}
\put(64,16){\line(1,0){9}}
\put(64,17){\line(1,0){9}}
\put(64,18){\line(1,0){9}}
\put(64,19){\line(1,0){9}}
\put(64,20){\line(1,0){9}}
\put(64,21){\line(1,0){9}}
\put(64,22){\line(1,0){9}}
\put(64,23){\line(1,0){9}}
\put(64,24){\line(1,0){9}}
\put(64,25){\line(1,0){9}}
\put(64,26){\line(1,0){9}}
\put(64,27){\line(1,0){9}}
\put(64,28){\line(1,0){9}}
\put(64,29){\line(1,0){9}}
\put(64,30){\line(1,0){9}}
\put(68.5,31){$\vdots$}

\put( 52,  28.72650){\line(1,0){9}}
\put( 52,  25.25990){\line(1,0){9}}
\put( 52,  22.92350){\line(1,0){9}}
\put( 52,  21.21150){\line(1,0){9}}
\put( 52,  19.82030){\line(1,0){9}}
\put( 52,  18.92580){\line(1,0){9}}
\put( 52,  17.68610){\line(1,0){9}}
\put( 52,  17.66200){\line(1,0){9}}
\put( 52,  16.51840){\line(1,0){9}}
\put( 52,  15.84030){\line(1,0){9}}
\put( 52,  14.88860){\line(1,0){9}}
\put( 52,  13.95424){\line(1,0){9}}
\put( 52,  12.97864){\line(1,0){9}}
\put( 52,  11.99176){\line(1,0){9}}
\put( 52,  10.99707){\line(1,0){9}}
\put( 52,   9.99911){\line(1,0){9}}
\put( 52,   8.99977){\line(1,0){9}}
\put( 52,   7.99995){\line(1,0){9}}
\put( 52,   6.99999){\line(1,0){9}}
\put( 52,   6.00000){\line(1,0){9}}
\put( 52,   5.00000){\line(1,0){9}}

\put(49,4){$\scriptscriptstyle{0}$}
\put(49,9){$\scriptscriptstyle{5}$}
\put(48.3,14){$\scriptscriptstyle{10}$}
\put(48.3,19){$\scriptscriptstyle{15}$}
\put(48.3,24){$\scriptscriptstyle{20}$}
\put(48.3,29){$\scriptscriptstyle{25}$}

\put( 32,   5.72035){\line(1,0){9}}
\put( 32,   7.10703){\line(1,0){9}}
\put( 32,   8.38750){\line(1,0){9}}
\put( 32,   9.56515){\line(1,0){9}}
\put( 32,  10.64288){\line(1,0){9}}
\put( 32,  11.62205){\line(1,0){9}}
\put( 32,  12.50795){\line(1,0){9}}
\put( 32,  13.27682){\line(1,0){9}}
\put( 32,  14.01840){\line(1,0){9}}
\put( 32,  14.43442){\line(1,0){9}}
\put( 32,  15.17428){\line(1,0){9}}
\put( 32,  15.32058){\line(1,0){9}}
\put( 32,  16.07323){\line(1,0){9}}
\put( 32,  16.69753){\line(1,0){9}}
\put( 32,  17.49758){\line(1,0){9}}
\put( 32,  18.37703){\line(1,0){9}}
\put( 32,  19.35725){\line(1,0){9}}
\put( 32,  20.43485){\line(1,0){9}}
\put( 32,  21.61250){\line(1,0){9}}
\put( 32,  22.89297){\line(1,0){9}}
\put( 32,  24.27966){\line(1,0){9}}
\put(29.9,29.4){\line(1,0){0.3}}
\put(29.9,24.6){\line(1,0){0.3}}
\put(29.9,19.8){\line(1,0){0.3}}
\put(29.9,15){\line(1,0){0.3}}
\put(29.9,10.2){\line(1,0){0.3}}
\put(29.9,5.4){\line(1,0){0.3}}
\put(29,29){$\scriptscriptstyle{2}$}
\put(29,24.1){$\scriptscriptstyle{0}$}
\put(28,19.3){$\scriptscriptstyle{-2}$}
\put(28,14.5){$\scriptscriptstyle{-4}$}
\put(28,9.7){$\scriptscriptstyle{-6}$}
\put(28,5){$\scriptscriptstyle{-8}$}

\put(36,2){$\scriptstyle \mathcal{H}_d$}
\put(68,2){$\scriptstyle{\hat H}$}
\put(56,2){$\scriptstyle{\bf H}_d$}
\end{picture}
\caption{\label{eigval}The eigenvalues of  $\hat H$  and the eigenvalues of $\mathcal{H}_d$ and  ${\bf H}_d$  in the case $d\!=\!21$.}
\end{figure}
\end{center}

\section{Finite oscillator obtained by using the  finite frame quantization}

The operator corresponding to the function
\begin{equation}
\begin{array}{l}
 f:\mathbb{R}^2\longrightarrow \mathbb{R}, \qquad f(\alpha ,\beta )=\frac{\alpha ^2\!+\!\beta ^2}{2}
 \end{array}
\end{equation}
through the {\em coherent state quantization} \cite{Ga}, namely,
\begin{equation}
\begin{array}{l}
A_f\!=\!\frac{1}{2\pi }\int_{\mathbb{R}^2}d\alpha \, d\beta \, 
\frac{\alpha ^2\!+\!\beta ^2}{2}\,|\alpha ,\beta \rangle \langle \alpha ,\beta |
\end{array}
\end{equation}
is the Hamiltonian of a translated harmonic oscillator \cite{Ga}
\begin{equation}
\begin{array}{l}
A_f \!=\!\sum\limits_{n=0}^\infty (n\!+\!1)\, |n\rangle \langle n|
=-\frac{1}{2}\, \frac{{\rm d}^2}{{\rm d}x^2}+\frac{1}{2}x^2+\frac{1}{2}=\hat H+\frac{1}{2}.
\end{array}
\end{equation}
The relation
\begin{equation}
\begin{array}{l}
 \hat H=-\frac{1}{2}+\frac{1}{2\pi }\int_{\mathbb{R}^2}d\alpha \, d\beta \, 
\frac{\alpha ^2\!+\!\beta ^2}{2}\,|\alpha ,\beta \rangle \langle \alpha ,\beta |
\end{array}
\end{equation}
can be regarded as an integral representation of the harmonic oscillator Hamiltonian. 

The linear operator corresponding to the function 
\begin{equation}
\begin{array}{l}
{\bf f}\!:\!\mathcal{R}_d^2\!\longrightarrow \!\mathbb{R}, \qquad  
{\bf f}(\alpha ,\beta )=\frac{\alpha ^2\!+\!\beta ^2}{2}
\end{array}
\end{equation}
through the {\em finite frame quantization}, namely,
\begin{equation}
\begin{array}{l}
{\bf A}_{\bf f}:l^2(\mathcal{R}_d)\longrightarrow l^2(\mathcal{R}_d),\qquad 
{\bf A}_{\bf f}=\frac{1}{d}\sum\limits_{(\alpha ,\beta )\in \mathcal{R}_d^2}\frac{\alpha ^2\!+\!\beta ^2}{2}\, |\alpha ,\beta \rangle _{\!\mbox{}_d}\, {}_{\mbox{}_d} \!\langle \alpha ,\beta |.
\end{array}
\end{equation}
can be regarded \cite{CGV} as a finite counterpart of $A_f$. The operator 
\begin{equation}
{\bf H}_d={\bf A}_{\bf f}- \frac{1}{2}
\end{equation}
is the Hamiltonian of a  finite oscillator. We prove that, in a certain sense, 
\begin{equation}
 \hat H\!=\!\lim_{d\rightarrow \infty }{\bf H}_d.
\end{equation}
The improper integral 
\[
\begin{array}{l}
A_f\!=\!\frac{1}{2\pi }\int_{\mathbb{R}^2}d\alpha \, d\beta \, \frac{\alpha ^2\!+\!\beta ^2}{2}\,|\alpha ,\beta \rangle \langle \alpha ,\beta |
\end{array}
\]
can be defined as the limit
\[
A_f\!=\!\lim_{d\rightarrow \infty }\int_{S_d}d\alpha \, d\beta \, \frac{\alpha ^2\!+\!\beta ^2}{2}\,|\alpha ,\beta \rangle \langle \alpha ,\beta |
\]
where $S_d$ is the square
\[
\begin{array}{l}
S_d=\left[-(s+\frac{1}{2})\sqrt{\delta }, (s+\frac{1}{2})\sqrt{\delta }\right] \times \left[-(s+\frac{1}{2})\sqrt{\delta }, (s+\frac{1}{2})\sqrt{\delta }\right] .
\end{array}
\]
By using a subdivison of  $S_d$ into $d^2$ squares of side $\sqrt{\delta }=\sqrt{2\pi /d}$, we can write
$A_f$ as a limit of Riemann sums
\[
\begin{array}{rl}
A_f & \!=\!\lim_{d\rightarrow \infty }\frac{1}{2\pi }\sum\limits_{(\alpha ,\beta )\in \mathcal{R}_d^2} \frac{2\pi }{d}\, \frac{\alpha ^2\!+\!\beta ^2}{2}\, |\alpha ,\beta \rangle \langle \alpha ,\beta |\\[5mm]
 & \!=\!\lim_{d\rightarrow \infty }\frac{1}{d}\sum\limits_{(\alpha ,\beta )\in \mathcal{R}_d^2} \frac{\alpha ^2\!+\!\beta ^2}{2}\, |\alpha ,\beta \rangle \langle \alpha ,\beta |.
\end{array}
\]
For any $\psi $ we have
\[
\begin{array}{rl}
|\alpha ,\beta \rangle \langle \alpha ,\beta |\psi \rangle & =|\alpha ,\beta \rangle \, \int_\mathbb{R}\overline{\mathcal{D}(\alpha ,\beta )\Psi_0(x)}\, \psi (x)\, dx\\[4mm]
& = \lim_{d\rightarrow \infty }|\alpha ,\beta \rangle \, \int _{-(s+\frac{1}{2})\sqrt{\delta }}^{ (s+\frac{1}{2})\sqrt{\delta }}\overline{\mathcal{D}(\alpha ,\beta )\Psi_0(x)}\, \psi (x)\, dx .
\end{array}
\]
By using a subdivision of the interval $\left[-(s\!+\!\frac{1}{2})\sqrt{\delta }, (s\!+\!\frac{1}{2})\sqrt{\delta }\right]$ 
into $d$ intervals of length $\sqrt{\delta }$ and the relation (\ref{approxcs1}) we get
\[
\begin{array}{rl}
|\alpha ,\beta \rangle \langle \alpha ,\beta |\psi \rangle & = \lim_{d\rightarrow \infty }|\alpha ,\beta \rangle \,\sum\limits_{u\in \mathcal{R}_d}\sqrt{\delta }\, \, \overline{\mathcal{D}(\alpha ,\beta )\Psi_0(u)}\, \psi (u)\\[3mm]
& \approx  \lim_{d\rightarrow \infty }\sqrt[4]{\delta }\, |\alpha ,\beta \rangle \,\sum\limits_{u\in \mathcal{R}_d}\overline{{\bf D}(\alpha ,\beta ){\bf g}(u)}\, \psi (u)\\[3mm]
& =\lim_{d\rightarrow \infty }|\alpha ,\beta \rangle _{\!\mbox{}_d}\, {}_{\mbox{}_d} \!\langle \alpha ,\beta |\psi \rangle .
\end{array}
\]
Therefore, in a certain sense, we have
\[
\begin{array}{rl}
A_f &  \!=\!\lim_{d\rightarrow \infty }\frac{1}{d}\sum\limits_{(\alpha ,\beta )\in \mathcal{R}_d^2} 
\frac{\alpha ^2\!+\!\beta ^2}{2}\, |\alpha ,\beta \rangle _{\!\mbox{}_d}\, {}_{\mbox{}_d} \!\langle \alpha ,\beta |.
\end{array}
\]

From the relation
\begin{equation}
\begin{array}{l}
\langle \varphi |{\bf A}_{\bf f}|\varphi \rangle =\frac{1}{d}\sum\limits_{(\alpha ,\beta )\in \mathcal{R}_d^2}\frac{\alpha ^2\!+\!\beta ^2}{2}\, 
|{}_{\mbox{}_d} \!\langle \alpha ,\beta |\varphi \rangle |^2\geq 0\qquad \mbox{for}\quad \varphi \!\in  \!l^2(\mathcal{R}_d)
\end{array}
\end{equation}
it follows that the eigenvalues of ${\bf A}_{\bf f}$ are non-negative.
The elements of the matrix $\left( \langle \varepsilon _n|{\bf H}_d|\varepsilon _m\rangle \right)_{-s\leq n,m\leq s}$ of ${\bf H_d}$ in the basis $\{ \varepsilon _n\}_{n=-s}^s$, namely, 
\begin{equation} \fl
\begin{array}{l}
 \langle \varepsilon _n|{\bf H}_d|\varepsilon _m\rangle \!= \!-\frac{1}{2}\delta _{nm}\!+\!
\frac{\pi }{d^2}\!\sum\limits_{a,b=-s}^s \!(a^2\!+\!b^2)\, 
{\rm e}^{\frac{2\pi {\rm i}}{d}b (n-m)} {\bf g}((n-a)\sqrt{\delta})\, {\bf g}((m-a)\sqrt{\delta})
\end{array}
\end{equation}
are real numbers described by periodic functions with respect to $n$ and $m$ and such that
\begin{equation}\label{propHd}
\langle \varepsilon _n|{\bf H}_d|\varepsilon _m\rangle =\langle \varepsilon _m|{\bf H}_d|\varepsilon _n\rangle =\langle \varepsilon _{-n}|{\bf H}_d|\varepsilon _{-m}\rangle .
\end{equation}
If $(v_{-s},v_{-s+1},...,v_{s-1}, v_s)$ is an eigenvector of ${\bf H}_d$ corresponding to the eigenvalue $\lambda $ then
\begin{equation}\label{eigHd}
 \sum_{m=-s}^s\langle \varepsilon _n|{\bf H}_d|\varepsilon _m\rangle \, v_m\!=\!\lambda \, v_n\qquad 
 \mbox{for any}\quad n\!\in \!\{ -s,-s\!+\!1,...,s\!-\!1,s\}.
\end{equation}
This relation being equivalent to
\[
 \sum_{m=-s}^s\langle \varepsilon _n|{\bf H}_d|\varepsilon _m\rangle \, \bar v_m=\lambda \, \bar v_n\qquad 
 \mbox{for any}\quad n\in \{ -s,-s+1,...,s-1,s\}
\]
we can choose a real eigenvector. But, by using (\ref{propHd}) the relation (\ref{eigHd}) can be written as
\[
 \sum_{m=-s}^s\langle \varepsilon _n|{\bf H}_d|\varepsilon _m\rangle \, v_{-m}=\lambda \, v_{-n}\qquad 
 \mbox{for any}\quad n\in \{ -s,-s+1,...,s-1,s\}.
\]
and this is possible only in the cases
\[
 (v_s,v_{s-1},...,v_{-s+1}, v_{-s})=\pm (v_{-s},v_{-s+1},...,v_{s-1}, v_s).
\]
Therefore any eigenfunction of ${\bf H}_d$ is either an even function or an odd one.
By using some results concerning the centrosymmetric matrices \cite{Mu}, one can prove that ${\bf H}_d$  admits $s$ even and $s\!+\!1$  odd eigenfunctions.

\begin{figure}[t]
\centering
\includegraphics[scale=0.7]{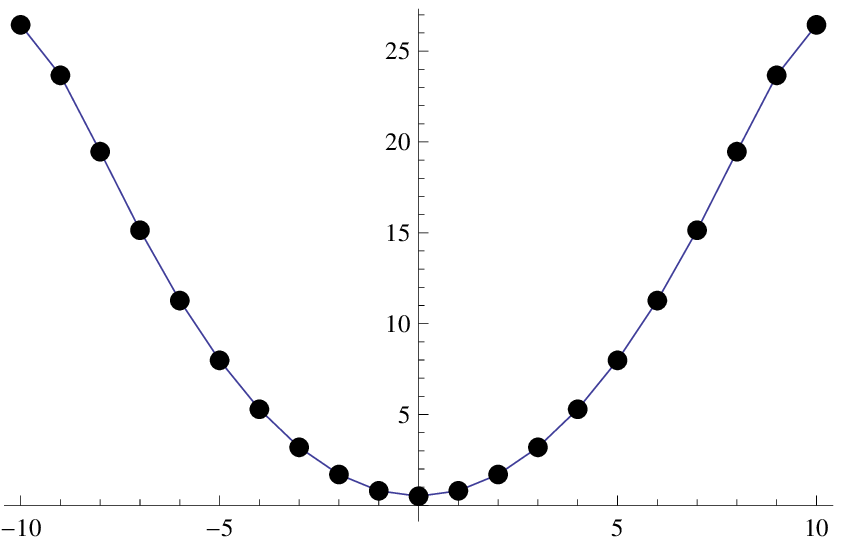}\qquad 
\includegraphics[scale=0.7]{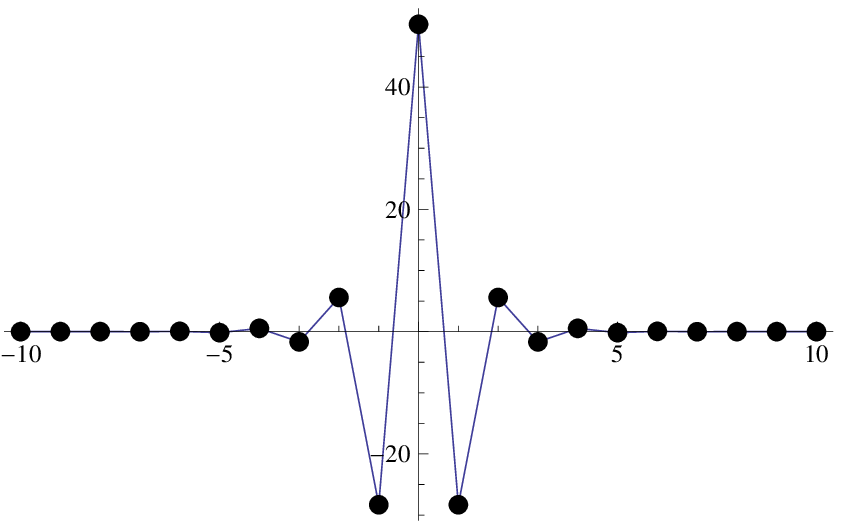}\\[2mm]
\caption{\label{conv} The convolution $\mathfrak{q}^2\!\ast \!{\bf g}^2$ (left) and its Fourier transform ${\bf F}[\mathfrak{q}^2\!\ast \!{\bf g}^2]$ (right).}
\end{figure}

The relation (\ref{normofg}) and Theorem 2 allows us to write the diagonal elements as
\begin{equation}\label{diag}
\begin{array}{l}
 \langle \varepsilon _n|{\bf H}_d|\varepsilon _n\rangle =-\frac{1}{2}+\frac{\pi }{d}\frac{s(s+1)}{3}+
\frac{\pi }{d}\sum\limits_{a=-s}^sa^2\, {\bf g}^2((n-a)\sqrt{\delta })
\end{array}
\end{equation}
and the non-diagonal elements as
\begin{equation}\label{nondiag}
\begin{array}{l}
 \langle \varepsilon _n|{\bf H}_d|\varepsilon _m\rangle =
\frac{1}{2}\, {\bf F}[\mathfrak{q}^2](n-m)\, \, 
{\bf F}[{\bf g}^2]((n\!-\!m)\sqrt{\delta }).
\end{array}
\end{equation}
Particularly, we have
\begin{equation}
\begin{array}{l}
 {\rm tr} \, {\bf H}_d=-\frac{d}{2}+2\pi \frac{s(s+1)}{3}.
\end{array}
\end{equation}
Numerically one can check (see Figure \ref{eigval}) that the eigenvalues of ${\bf H}_d$ are distinct and have the tendency to become
$\frac{1}{2}$, $\frac{1}{2}\!+\!1$, $\frac{1}{2}\!+\!2$, ..., $\frac{1}{2}\!+\!d\!-\!1$,  for large $d$. Since
\begin{equation}
\begin{array}{l}
 \sum\limits_{n=0}^{d-1}\left( \frac{1}{2}\!+\!n\right) =\frac{d^2}{2}
\end{array}
\end{equation}
one can remark that
\begin{equation}\label{limtr}
\lim_{d\rightarrow \infty }\frac{ {\rm tr} \, {\bf H}_d}{\sum_{n=0}^{d-1}\left( \frac{1}{2}\!+\!n\right)}=\frac{\pi }{3}\approx 1.0472.
\end{equation}
\mbox{}\\[1mm]
{\bf Theorem 3}. {\it The matrix $\left(\langle \varepsilon _n|{\bf A}_{\bf f}|\varepsilon _m\rangle \right)_{-s\leq n,m\leq s}$ of ${\bf A}_{\bf f}\!=\!{\bf H}_d\!+\!\frac{1}{2}$ in the basis $\{ \varepsilon _n\}_{n=-s}^s$ is
\begin{equation}
{\bf A}_{\bf f}\!=\!\left(
\begin{array}{lllllllllll}
\omega _s & \tau _1 & \tau _2 & \dots & \tau _{s-1}& \tau _s & \tau _s & \dots & \tau _2 & \tau _1\\
\tau _1  & \omega _{s-1} & \tau _1  &  \dots  & \tau _{s-2}& \tau _{s-1} & \tau _s & \dots & \tau _3 & \tau _2 \\
\tau _2  & \tau _1 & \omega _{s-2} & \dots  &\tau _{s-3}& \tau _{s-2} & \tau _{s-1} & \dots & \tau _4 & \tau _3 \\
\vdots & \vdots & \vdots & \ddots & \vdots & \vdots & \vdots & \vdots  & \vdots &\vdots \\
\tau _{s-1}  & \tau _{s-2} & \tau _{s-3} & \dots  & \omega _1& \tau _1 & \tau _2 & \dots & \tau _s & \tau _s \\
\tau _s  & \tau _{s-1} & \tau _{s-2} & \dots  & \tau _1& \omega _0 & \tau _1 & \dots & \tau _{s-1} & \tau _s \\
\tau _s  & \tau _s & \tau _{s-1} & \dots  & \tau _2& \tau _1 & \omega _1 & \dots & \tau _{s-2} & \tau _{s-1} \\
\vdots & \vdots & \vdots & \vdots & \vdots & \vdots & \vdots & \ddots & \vdots &\vdots  \\
\tau _2 & \tau _3  & \tau _4 & \dots  &\tau _s & \tau _{s-1} &\tau _{s-2}& \dots & \omega _{s-1} & \tau _1 \\
\tau _1 & \tau _2  & \tau _3 & \dots  & \tau _s& \tau _s & \tau _{s-1}& \dots & \tau _1 & \omega _s 
\end{array}\right)
\end{equation}
where}
\begin{equation}\label{tau-omega}
\begin{array}{l}
\tau _k\!=\!\frac{1}{2\sqrt{d}}\, {\bf F}[\mathfrak{q}^2\!\ast \!{\bf g}^2](k\sqrt{\delta})\qquad and\qquad 
\omega _k\!=\!\tau _0\!+\!\frac{1}{2}  (\mathfrak{q}^2\!\ast \!{\bf g}^2)(k\sqrt{\delta}).
\end{array}
\end{equation}
{\bf Proof}. The diagonal elements depend only on $|n|$ and the non-diagonal elements only on $|n\!-\!m|$ (see Figure \ref{conv}).
The relations (\ref{tau-omega}) follow from (\ref{diag}), (\ref{nondiag}) and
\[ \fl
\begin{array}{l}
\tau _0\!=\!\frac{1}{2d}\, \sum\limits_{n=-s}^s(\mathfrak{q}^2\!\ast \!{\bf g}^2)(n\sqrt{\delta})
\!=\!\frac{\pi }{d^2}\, \sum\limits_{n,k=-s}^s k^2\, {\bf g}^2((n-k)\sqrt{\delta})
\!=\!\frac{\pi }{d^2}\, \sum\limits_{k=-s}^s k^2=\frac{\pi }{d}\frac{s(s+1)}{3}.  \qquad \opensquare 
\end{array}
\]

The circulant matrix 
\begin{equation}
 T_d\!=\!\left( \tau _{n-m}\right)_{-s\leq n,m\leq s}=\left(
\begin{array}{lllllll}
\tau _0 & \tau _1 & \tau _2 & \dots & \tau _2 & \tau _1\\
\tau _1  & \tau _0 & \tau _1  &  \dots   & \tau _3 & \tau _2 \\
\tau _2  & \tau _1 &\tau _0 & \dots   & \tau _4 & \tau _3 \\
\vdots & \vdots & \vdots & \ddots  & \vdots &\vdots \\
\tau _2 & \tau _3  & \tau _4 & \dots  & \tau _0 & \tau _1 \\
\tau _1 & \tau _2  & \tau _3 & \dots   & \tau _1 & \tau _0
\end{array}\right)
\end{equation}
has the eigenvalues \cite{Gray}
\begin{equation}
\begin{array}{l}
\sum\limits_{n=-s}^s\tau _n\, {\rm e}^{\frac{2\pi {\rm i}}{d}kn}=\frac{1}{2}  (\mathfrak{q}^2\!\ast \!{\bf g}^2)(k\sqrt{\delta})
\qquad \mbox{with}\quad 
k\!\in \!\{ -s,-s\!+\!1,...,s\!-\!1,s\}
\end{array}
\end{equation}
and satisfies the relation
\begin{equation}
T_d={\bf F}^+\mathfrak{D}_{\!f}{\bf F}
\end{equation}
where $\mathfrak{D}_{\!f}$ is the diagonal matrix
\begin{equation}\fl
\begin{array}{l}
\mathfrak{D}_{\!f}\!=\!{\rm diag} \left( \frac{1}{2}  (\mathfrak{q}^2\!\ast \!{\bf g}^2)(-s\sqrt{\delta}) , \frac{1}{2}  (\mathfrak{q}^2\!\ast \!{\bf g}^2)((-s\!+\!1)\sqrt{\delta}),..., \frac{1}{2}  (\mathfrak{q}^2\!\ast \!{\bf g}^2)(s\sqrt{\delta}) \right).
\end{array}
\end{equation}
The matrix  of ${\bf H}_d$ in the basis $\{ \varepsilon _n\}_{n=-s}^s$ can be written as
\begin{equation}
\begin{array}{l}
{\bf H}_d=-\frac{1}{2}+\mathfrak{D}_{\!f}+{\bf F}^+\mathfrak{D}_{\!f}{\bf F}.
\end{array}
\end{equation}
Up to a translation, it is similar to the matrix of  $\frac{1}{2}({\bf P}^2\!+\!{\bf Q}^2)$  which can be written as 
\begin{equation}
\begin{array}{l}
\frac{1}{2}({\bf P}^2\!+\!{\bf Q}^2)=\mathfrak{D}+{\bf F}^+\mathfrak{D}{\bf F}.
\end{array}
\end{equation}
by using the diagonal matrix
 \begin{equation}
\begin{array}{l}
\mathfrak{D}\!=\!{\rm diag} \left( \frac{1}{2}  \mathfrak{q}^2(-s\sqrt{\delta}) , \frac{1}{2}  \mathfrak{q}^2((-s\!+\!1)\sqrt{\delta}),..., \frac{1}{2}  \mathfrak{q}^2(s\sqrt{\delta}) \right).
\end{array}
\end{equation}
Since ${\bf F}|\alpha ,\beta \rangle _{\!\mbox{}_d}=|-\beta , \alpha \rangle _{\!\mbox{}_d}$, 
the mean value of ${\bf H}_d$ in the state $|\alpha ,\beta \rangle _{\!\mbox{}_d}$ is
\begin{equation} 
{}_{\mbox{}_d} \!\langle \alpha ,\beta  |{\bf H}_d|\alpha ,\beta \rangle _{\!\mbox{}_d}
\!=\!-\frac{1}{2}\!+\!\frac{1}{2} \sum\limits_{u \in \mathcal{R}_d} (\mathfrak{q}^2\!\ast \!{\bf g}^2)(u)
\left({\bf g}^2(u\!-\!\alpha)\!+\!{\bf g}^2(u\!-\!\beta)\right).
\end{equation}

The circulant matrix with the equidistant eigenvalues $1$, $2$, $3$, ...,  $d$ is 
\begin{equation}
C_d\!={\bf F}^+{\rm diag}(1,2,...,d){\bf F}=\!\left( c_{n-m}\right)_{-s\leq n,m\leq s}
\end{equation}
where
\begin{equation}
\begin{array}{l}
c_k=\frac{1}{d}\sum\limits_{n=-s}^sn\, {\rm e}^{\frac{2\pi {\rm i}}{d}kn}=\left\{
\begin{array}{cll}
\frac{d+1}{2} & \mbox{for} & k \in d\mathbb{Z}\\[2mm]
\frac{{\rm e}^{\frac{2\pi {\rm i}}{d}k}}{{\rm e}^{\frac{2\pi {\rm i}}{d}k}-1}& \mbox{for} & k\not\in d\mathbb{Z}.
\end{array} \right. 
\end{array}
\end{equation}
Wielandt-Hoffman theorem \cite{Gray} admits as a direct consequence the following result.\\[5mm]
{\bf Theorem 4}.\cite{Gray} {\it Given two Hermitian matrices $A=(a_{ij})_{1\leq i,j\leq d}$ and $B=(b_{ij})_{1\leq i,j\leq d}$ with eigenvalues $\alpha _n$ and $\beta _n$ in nondecreasing order, respectively, then }
\begin{equation}
\frac{1}{d}\sum_{k=1}^d|\alpha _k-\beta _k|\leq ||A-B||.
\end{equation}
In view of this theorem,  the  eigenvalues $\lambda _1$, $\lambda _2$, ... , $\lambda _{d}$ of ${\bf A}_{\bf f}$
considered in a nondecreasing order satisfy the relation
\begin{equation}
\begin{array}{l}
\frac{1}{d}\sum\limits_{n=1}^{d}|n-\lambda _n|\leq \left(\sum\limits_{k=1}^{d-1}|\tau _k\!-\!c_k|^2+\frac{1}{d}\sum\limits_{k=-s}^s|\omega _k-c_0|^2\right)^{\frac{1}{2}}.
\end{array}
\end{equation}
Numerically one can check (see Figure \ref{eigval}) that $\lambda _1$, $\lambda _2$, ... , $\lambda _{d}$ have the tendency to become
$1$, $2$, ... , $d$,  for large $d$. \\[5mm]
{\bf Theorem 5}. {\it The Hamiltonian ${\bf H}_d$ is Fourier invariant}
\begin{equation}
{\bf F}{\bf H}_d={\bf H}_d{\bf F}.
\end{equation}
{\bf Proof}. By using the relation  (\ref{Fouriercs}) we get
\[
\begin{array}{rl}
{\bf F}{\bf H}_d{\bf F}^+ & \!\!\!\!=-\frac{1}{2}+\frac{1}{d}\sum\limits_{(\alpha ,\beta )\in \mathcal{R}_d^2}\frac{\alpha ^2\!+\!\beta ^2}{2}\,{\bf F} |\alpha ,\beta \rangle _{\!\mbox{}_d}\, {}_{\mbox{}_d} \!\langle \alpha ,\beta |{\bf F}^+\\[3mm]
& \!\!\!\!=-\frac{1}{2}+\frac{1}{d}\sum\limits_{(\alpha ,\beta )\in \mathcal{R}_d^2}\frac{\alpha ^2\!+\!\beta ^2}{2}\, |-\beta , \alpha  \rangle _{\!\mbox{}_d}\, {}_{\mbox{}_d} \!\langle -\beta ,\alpha |={\bf H}_d. \qquad \opensquare 
\end{array}
\]

\begin{figure}[h]
\centering
\includegraphics[scale=0.8]{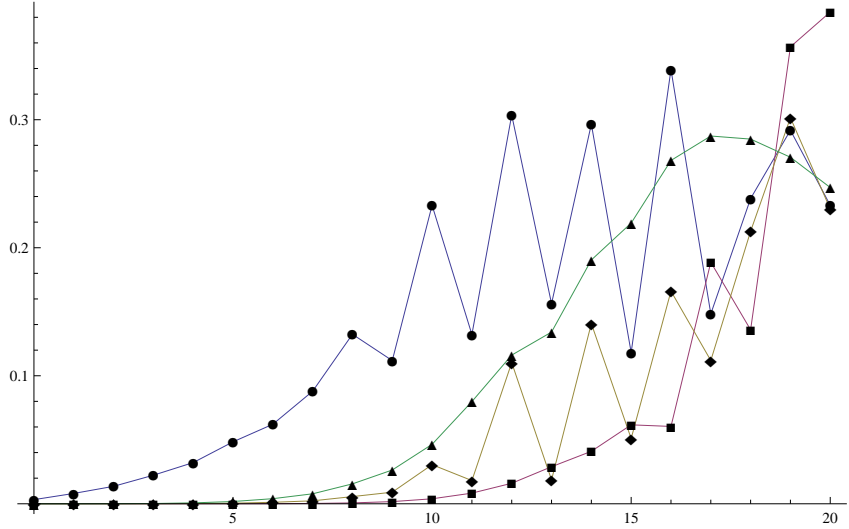}
\caption{\label{Deltafig} The values of $\Delta (m)$ (rhombs), $\Delta _H(m)$ (bullets), $\Delta _M(m)$ (squares)  
 and  $\Delta _R(m)$ (triangles) in the case $d\!=\!21$. }
\end{figure}

One can check that the normalized eigenfunctions ${\bf f}_0,\, {\bf f}_1,...,{\bf f}_{d-1}$ of ${\bf H}_d$, considered in the increasing order of the 
number of sign alternations, satisfy the relation
\begin{equation}
 {\bf F}\, {\bf f}_m=(-{\rm i})^m\, {\bf f}_m
\end{equation}
and approximate the Hermite-Gaussian functions $\Psi _m$ better than the Harper functions ${\bf h}_m$. 
By denoting 
\begin{equation}\label{deltaH}
\begin{array}{l}
\Delta (m)=\max\limits_{u\in \mathcal{R}_d} |{\bf f}_m (u)-\sqrt[4]{\delta }\, \Psi _m(u)|\\[2mm]
\Delta _H(m)=\max\limits_{u\in \mathcal{R}_d} |{\bf h}_m (u)-\sqrt[4]{\delta }\, \Psi _m(u)|
\end{array}
\end{equation}
we have (see Figure \ref{Deltafig})
\begin{equation}
\Delta (m)<\Delta _H(m)\qquad \mbox{for almost all} \quad  m\!\in \!\{ 0,1,2,...,d\!-\!1\}.
\end{equation}

\section{A finite counterpart of the raising operator}

The operator corresponding through the coherent state quantization to the function 
\begin{equation}
\begin{array}{l}
 \mathbb{R}^2\longrightarrow \mathbb{C}\, :\,(\alpha ,\beta )\mapsto \frac{\alpha \!-\!{\rm i}\beta }{\sqrt{2}}
 \end{array}
\end{equation}
is the usual raising operator \cite{Ga}
\begin{equation}
\begin{array}{l}
\frac{1}{2\pi }\int_{\mathbb{R}^2}d\alpha \, d\beta \, 
\frac{\alpha \!-\!{\rm i}\beta }{\sqrt{2}}\,|\alpha ,\beta \rangle \langle \alpha ,\beta |=a^+
\end{array}
\end{equation}
satisfying the relations
\begin{equation}
a^+\Psi _n=\sqrt{n\!+\!1}\, \Psi _{n+1}\qquad \mbox{and}\qquad \Psi _n=\frac{1}{\sqrt{n!}}(a^+)^n\Psi _0.
\end{equation}
By using the finite frame quantization  we define a finite counterpart of $a^+$, namely,
\begin{equation} 
\begin{array}{l}
{\bf a}_d^+:l^2(\mathcal{R}_d)\longrightarrow l^2(\mathcal{R}_d),\qquad 
{\bf a}_d^+=\frac{1}{d}\sum\limits_{(\alpha ,\beta )\in \mathcal{R}_d^2}\frac{\alpha \!-\!{\rm i}\beta }{\sqrt{2}}\, |\alpha ,\beta \rangle _{\!\mbox{}_d}\, {}_{\mbox{}_d} \!\langle \alpha ,\beta |.
\end{array}
\end{equation}
The elements of the matrix $\left( \langle \varepsilon _n|{\bf a}_d^+|\varepsilon _m\rangle \right)_{-s\leq n,m\leq s}$ of ${\bf a}_d^+$ in the basis $\{ \varepsilon _n\}_{n=-s}^s$, namely,
\begin{equation}\fl
\begin{array}{l}
 \langle \varepsilon _n|{\bf a}_d^+|\varepsilon _m\rangle \!= \!
\frac{1}{d}\sqrt{\frac{\pi }{d}}\!\sum\limits_{a,b=-s}^s \!(a\!-\!{\rm i}b)\, 
{\rm e}^{\frac{2\pi {\rm i}}{d}b (n-m)}{\bf g}((n-a)\sqrt{\delta})\, {\bf g}((m-a)\sqrt{\delta})
\end{array}
\end{equation}
are real numbers described by periodic functions with respect to $n$ and $m$ and such that
\begin{equation}
\langle \varepsilon _n|{\bf a}_d^+|\varepsilon _m\rangle =-\langle \varepsilon _{-n}|{\bf a}_d^+|\varepsilon _{-m}\rangle .
\end{equation}
The functions $\tilde {\bf f}_0$, $\tilde {\bf f}_1$, ... , $\tilde {\bf f}_{d-1}$ defined by the relation
\begin{equation}
\tilde {\bf f}_n=\frac{1}{\sqrt{n!}}({\bf a}_d^+)^n{\bf g}
\end{equation}
can be regarded as a finite counterpart of the Hermite-Gaussian functions.
They satisfy the recurrence relation 
\begin{equation}
{\bf a}_d^+\tilde {\bf f}_n=\sqrt{n\!+\!1}\, \tilde {\bf f} _{n+1}.
\end{equation}

\section{A discrete fractional Fourier transform}

The continuous Fourier transform satisfies the relation
\begin{equation}
\begin{array}{rl}
\mathcal{F}[\psi ](x) & =\mathcal{F}\left[ \sum\limits_{m=0}^\infty \langle \Psi _m,\psi \rangle \, \Psi _m \right](x)
= \sum\limits_{m=0}^\infty \langle \Psi _m,\psi \rangle \, \mathcal{F}[\Psi _m ](x)\\[3mm]
 & =\sum\limits_{m=0}^\infty (-{\rm i})^m\, \Psi _m(x)\, \int_{-\infty }^\infty \Psi _m(x')\, \psi (x')\, dx'\\[3mm]
& = \int_{-\infty }^\infty \left[ \sum\limits_{m=0}^\infty {\rm e}^{-\frac{\pi {\rm i}}{2}m}\, \Psi _m(x)\, \Psi _m(x')\right] 
\psi (x')\, dx'
\end{array}
\end{equation}
and the usual $\alpha $th-order continuous {\em fractional Fourier transform} is defined as 
\begin{equation}
\mathcal{F}^\alpha [\psi ](x)=\int_{-\infty }^\infty \left[ \sum\limits_{m=0}^\infty {\rm e}^{-\frac{\pi {\rm i}}{2}m\alpha }\, 
\Psi _m(x)\, \Psi _m(x')\right] \psi (x')\, dx'.
\end{equation}
The finite Fourier transform $l^2(\mathcal{R}_d)\longrightarrow l^2(\mathcal{R}_d):\, \varphi \mapsto {\bf F}[\varphi ]$
satisfies the relation 
\begin{equation}
\begin{array}{rl}
{\bf F}[\varphi ](u) & ={\bf F}\left[ \sum\limits_{m=0}^{d-1} \langle {\bf h}_m,\varphi \rangle \, {\bf h}_m \right](u)
= \sum\limits_{m=0}^{d-1} \langle {\bf h}_m,\varphi \rangle \, {\bf F}[{\bf h}_m ](u)\\[3mm]
 & =\sum\limits_{m=0}^{d-1} (-{\rm i})^m\, {\bf h}_m(x)\, \sum\limits_{v\in \mathcal{R}_d}{\bf h}_m(v)\, \varphi (v)\\[3mm]
& = \sum\limits_{v\in \mathcal{R}_d}\left[ \sum\limits_{m=0}^{d-1} 
{\rm e}^{-\frac{\pi {\rm i}}{2}m}\, {\bf h}_m(u)\, {\bf h}_m(v)\right] \varphi (v).
\end{array}
\end{equation}
and the transformation
\[
l^2(\mathcal{R}_d)\longrightarrow l^2(\mathcal{R}_d):\, \varphi \mapsto {\bf F}_H^\alpha [\varphi ]
\]
where
\begin{equation}\label{Harperdef}
 {\bf F}_H^\alpha [\varphi ](u)=\sum\limits_{v\in \mathcal{R}_d}\left[ \sum\limits_{m=0}^{d-1} 
{\rm e}^{-\frac{\pi {\rm i}}{2}m\alpha }\, {\bf h}_m(u)\, {\bf h}_m(v)\right] \varphi (v)
\end{equation}
is called a {\em discrete fractional Fourier transform}. It is a unitary transformation
\begin{equation}
{\bf F}_H^\alpha \, ({\bf F}_H^\alpha )^+=({\bf F}_H^\alpha )^+\, {\bf F}_H^\alpha ={\bf I}
\end{equation}
and satisfies the relations
\begin{equation}
{\bf F}_H^0={\bf I},\qquad {\bf F}_H^1={\bf F}\qquad \mbox{and}\qquad {\bf F}_H^\alpha \, 
{\bf F}_H^\beta ={\bf F}_H^{\alpha  +\beta}.
\end{equation}

\section{An alternative discrete fractional Fourier transform}

The finite Fourier transform $l^2(\mathcal{R}_d)\longrightarrow l^2(\mathcal{R}_d):\, \varphi \mapsto {\bf F}[\varphi ]$
satisfies the relation 
\begin{equation}
\begin{array}{l}
{\bf F}[\varphi ](u) = \sum\limits_{v\in \mathcal{R}_d}\left[ \sum\limits_{m=0}^{d-1} 
{\rm e}^{-\frac{\pi {\rm i}}{2}m}\, {\bf f}_m(u)\, {\bf f}_m(v)\right] \varphi (v)
\end{array}
\end{equation}
and the transformation
\[
l^2(\mathcal{R}_d)\longrightarrow l^2(\mathcal{R}_d):\, \varphi \mapsto {\bf F}^\alpha [\varphi ]
\]
where
\begin{equation}\label{ourdef}
 {\bf F}^\alpha [\varphi ](u)=\sum\limits_{v\in \mathcal{R}_d}\left[ \sum\limits_{m=0}^{d-1} 
{\rm e}^{-\frac{\pi {\rm i}}{2}m\alpha }\, {\bf f}_m(u)\, {\bf f}_m(v)\right] \varphi (v)
\end{equation}
is also a {\em discrete fractional Fourier transform}. It is a unitary transformation
\begin{equation}
{\bf F}^\alpha \, ({\bf F}^\alpha )^+=({\bf F}^\alpha )^+\, {\bf F}^\alpha ={\bf I}
\end{equation}
and satisfies the relations
\begin{equation}
{\bf F}^0={\bf I},\qquad {\bf F}^1={\bf F}\qquad \mbox{and}\qquad {\bf F}^\alpha \, 
{\bf F}^\beta ={\bf F}^{\alpha  +\beta}.
\end{equation}

\begin{figure}[t]
\centering
\includegraphics[scale=0.7]{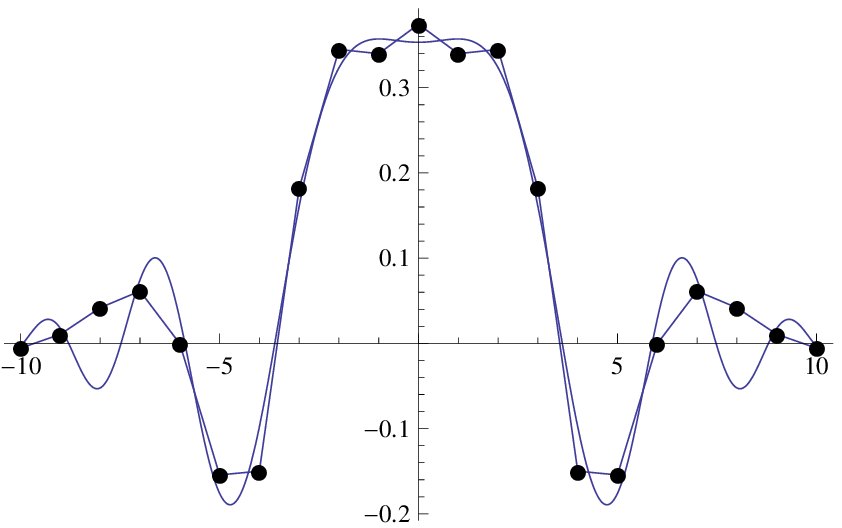}\qquad 
\includegraphics[scale=0.7]{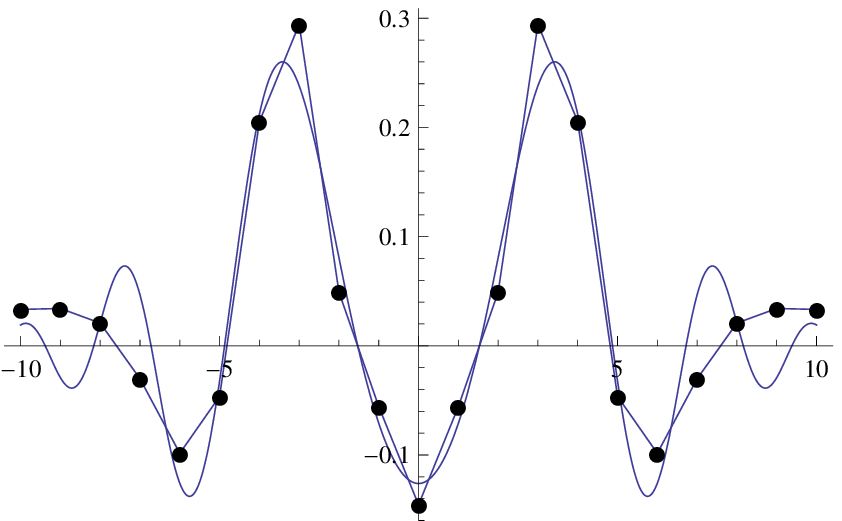}\\[2mm]
\includegraphics[scale=0.7]{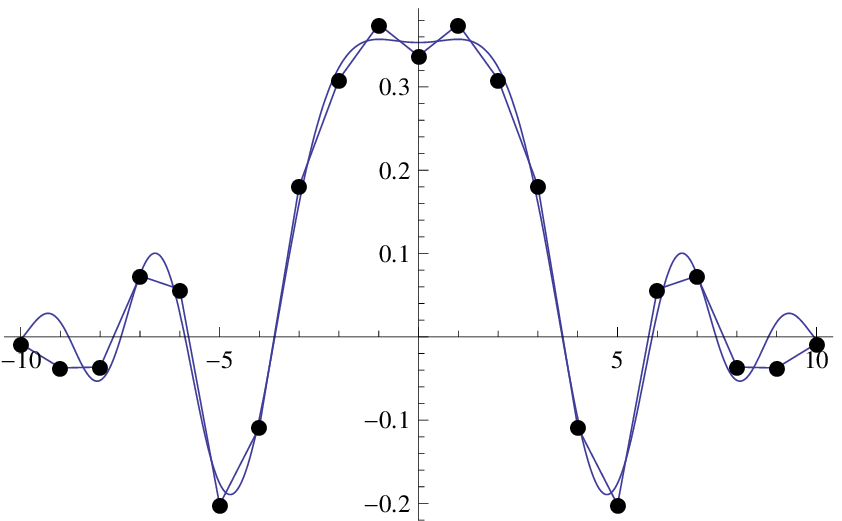}\qquad 
\includegraphics[scale=0.7]{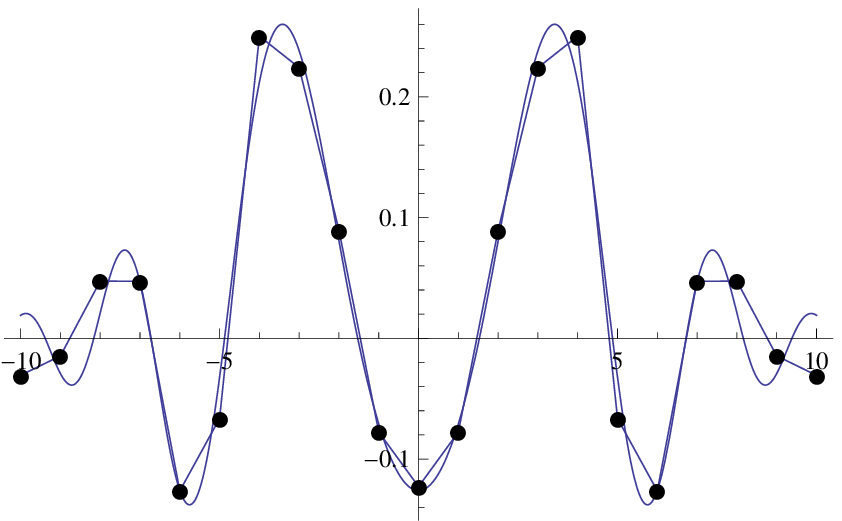}
\caption{\label{gauss05} The real part (left hand side) and the imaginary part (right hand side) of the $0.5$-th order fractional 
Fourier transform of the Gaussian $g_{10}$ and of its discrete counterpart $\mathfrak{g}_{10}$ computed by using the 
definition based on the Harper functions (first line) and the proposed definition (second line). }
\end{figure}  

\begin{figure}[t]
\centering
\includegraphics[scale=0.7]{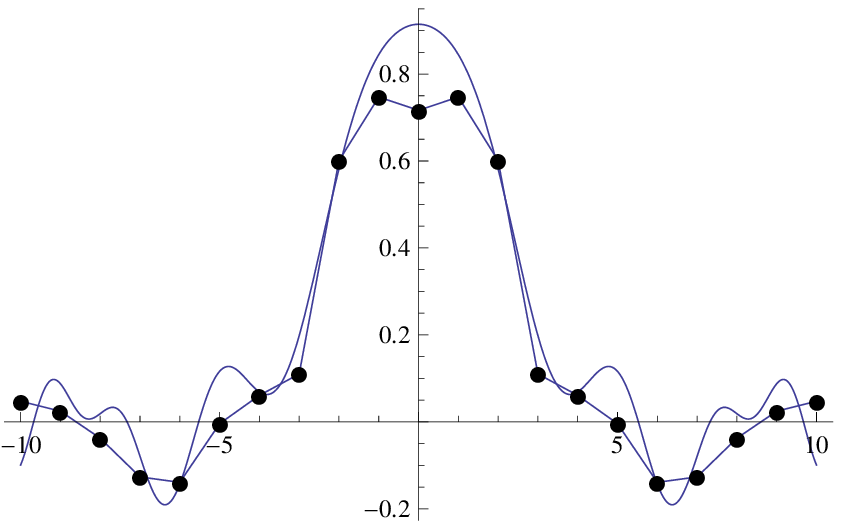}\qquad 
\includegraphics[scale=0.7]{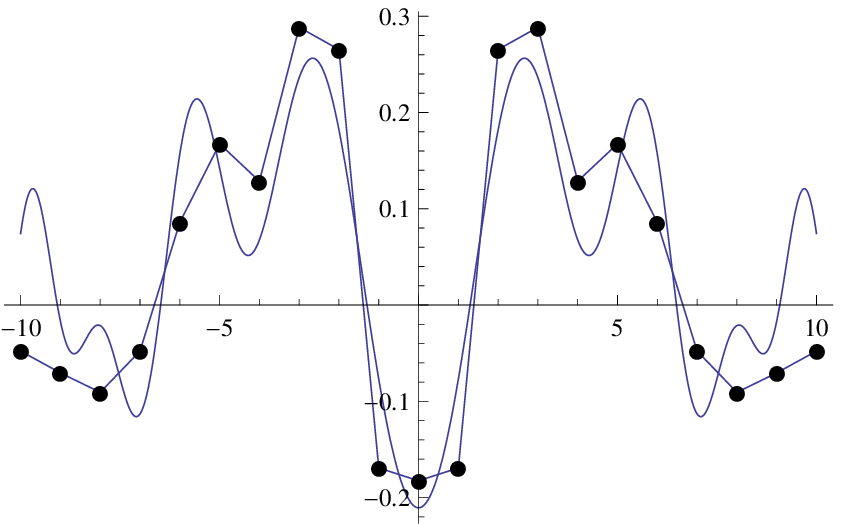}\\[2mm]
\includegraphics[scale=0.7]{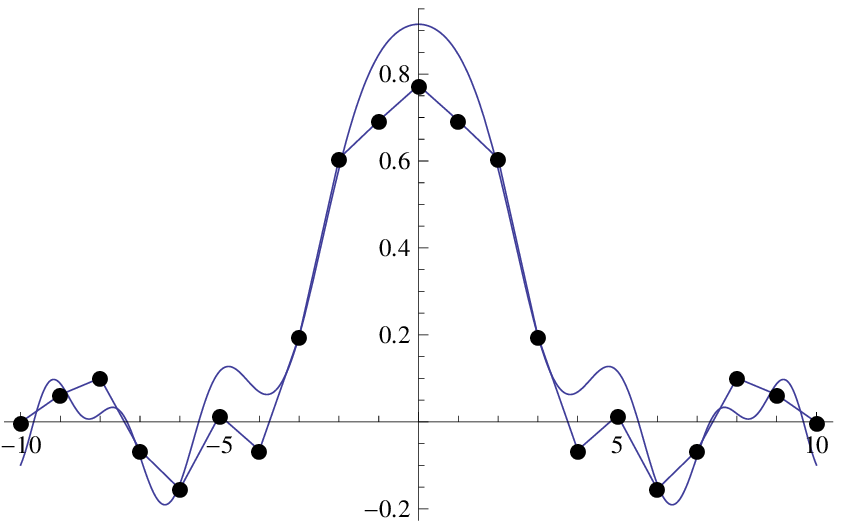}\qquad 
\includegraphics[scale=0.7]{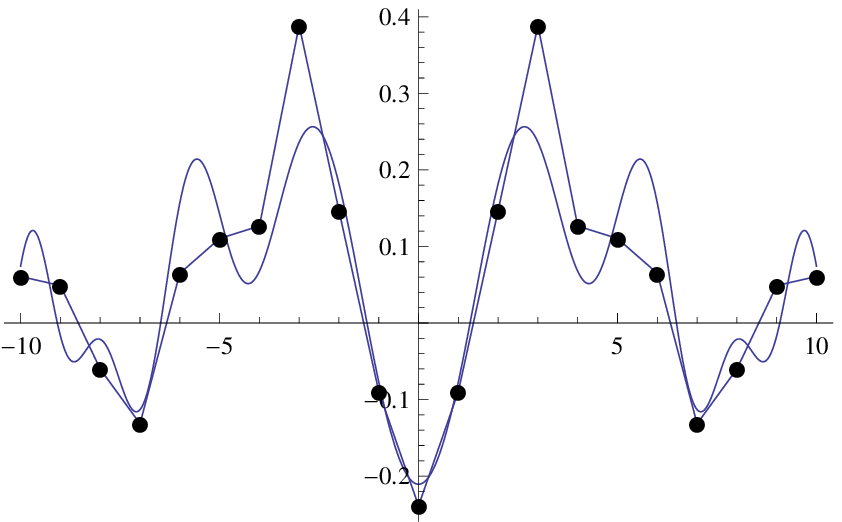}
\caption{\label{signal05} The real part (left hand side) and the imaginary part (right hand side) of the $0.5$-th order fractional 
Fourier transform of the rectangular function (\ref{csignal}) and of its discrete counterpart (\ref{dsignal}) computed by using the 
definition based on the Harper functions (first line) and the proposed definition (second line). }
\end{figure}

The discrete fractional Fourier transform computed by using our definition (\ref{ourdef}), generally,  approximates 
the corresponding continuous fractional Fourier transform better than the discrete fractional transformation 
computed by using the definition (\ref{Harperdef}) based on the Harper functions. 
For example, in the case $d\!=\!21$, the discrete counterpart of  the Gaussian
\begin{equation}
g_{10} :\mathbb{R}\longrightarrow \mathbb{R}, \qquad g_{10} (x)={\rm e}^{-5 x^2}
\end{equation}
is the periodic function (see Figure \ref{gaussians})
\begin{equation}
\mathfrak{g}_{10}  :\mathbb{Z}\sqrt{\delta }\longrightarrow \mathbb{R}, \qquad 
\mathfrak{g}_{10}  (n\sqrt{\delta })=\sum_{\ell =-\infty }^\infty {\rm e}^{-\frac{10 \pi }{21}(21\, \ell +n)^2}.
\end{equation}
In Figure \ref{gauss05} we present the graphs of the real and imaginary part of $\mathcal{F}^\alpha [g_{10}]$
 superposed on the corresponding graphs of ${\bf F}_H^\alpha [\mathfrak{g}_{10}]$ (first line)
and ${\bf F}^\alpha [\mathfrak{g}_{10}]$ (second line). In Figure \ref{signal05} we compare the $0.5$-th order fractional 
Fourier transforms of the rectangular function
\begin{equation}\label{csignal}
 \psi :\mathbb{R}\longrightarrow \mathbb{R},\qquad \psi (x)=\left\{ 
 \begin{array}{lll}
 1 & \mbox{for} & x\in [-\sqrt{\delta }, \sqrt{\delta }]\\
 0 & \mbox{for} & x\not\in [-\sqrt{\delta } , \sqrt{\delta }]  
 \end{array}\right.
\end{equation}
and of its discrete counterpart
\begin{equation}\label{dsignal}
 \varphi :\mathcal{R}_{21}\longrightarrow \mathbb{R},\qquad \varphi (x)=\left\{ 
 \begin{array}{lll}
 1 & \mbox{for} & x\in \{-\sqrt{\delta },0, \sqrt{\delta }\}\\
 0 & \mbox{for} & x\not\in \{-\sqrt{\delta },0, \sqrt{\delta }\}  
 \end{array}\right.
\end{equation}
computed by using the definition based on the Harper functions and our definition.

\section{Concluding remarks}
 In the case of a particle moving along a line, the momentum operator admits the differential representation 
\begin{equation}
\hat p=-{\rm i} \frac{\rm d}{{\rm d}q}
\end{equation}
as well as the integral representation
\begin{equation}
 (\hat p \psi )(q)=\frac{1}{2\pi }\int_{\mathbb{R}^2}x\, {\rm e}^{{\rm i}xy}\, \psi (q-y) dx\, dy
\end{equation}
equivalent to $\hat p =\mathcal{F}^+\hat q\mathcal{F}$.
In the case of a quantum system with finite-dimensional Hilbert space, the definition of the momentum operator is obtained
by starting from the integral representation, and not as finite difference operator.

The Hamiltonian of the harmonic oscillator admits the differential representation 
\begin{equation}
\hat H=-\frac{1}{2}\, \frac{{\rm d}^2}{{\rm d}x^2}+\frac{1}{2}x^2
\end{equation}
as well as the integral representation 
\begin{equation}\label{intrep}
\begin{array}{l}
 \hat H=-\frac{1}{2}+\frac{1}{2\pi }\int_{\mathbb{R}^2}d\alpha \, d\beta \, 
\frac{\alpha ^2\!+\!\beta ^2}{2}\,|\alpha ,\beta \rangle \langle \alpha ,\beta |.
\end{array}
\end{equation}
A finite counterpart is usually obtained by using a difference operator instead of the differential operator.
We think that, generally, the integral representations behave better than the differential representations 
when we have to define finite versions. The finite frame quantization is a finite counterpart of the
coherent state quantization, and we have used it in order to define a finite oscillator by starting from the
 integral representation (\ref{intrep}). We belive that it approximates the harmonic oscillator better than
the finite version based on the use of finite difference operators (see relation (\ref{limtr}), Figure \ref{eigval} and Figure \ref{Deltafig}).

There exists several systems of functions which can  be regarded as a finite counterpart of the Hermite-Gauss functions. Among them there are
the Mehta functions 
\begin{equation}
\begin{array}{l}
\Phi _m(n)=\sum\limits_{\ell =-\infty }^\infty \Psi _m \left((\ell d +n)\sqrt{\delta} \right)
\end{array}
\end{equation}
the Harper functions $\{ {\bf h}_m\}$, and our systems of functions $\{ {\bf f}_m\}$ and $\{ \tilde {\bf f}_m\}$.
In Figure \ref{Deltafig} these systems are compared with the Hermite-Gauss functions by using the notations (\ref{deltaH}) and
\begin{equation}
\begin{array}{l}
\Delta _M(m)=\max\limits_{u\in \mathcal{R}_d} |\Phi_m (u)-\sqrt[4]{\delta }\, \Psi _m(u)|\\[2mm]
\Delta _R(m)=\max\limits_{u\in \mathcal{R}_d} |\tilde{\bf f}_m (u)-\sqrt[4]{\delta }\, \Psi _m(u)|.
\end{array}
\end{equation}
\section*{References}

\end{document}